\documentclass[fleqn,twoside,onecolumn,nofootinbib,showkeys]{revtex4}
\usepackage{graphicx}
\usepackage{bm}
\begin{document}

\begin{center}

{\large \bf NON-SMOOTH CHEMICAL FREEZE-OUT 
AND APPARENT WIDTH OF WIDE RESONANCES AND QUARK GLUON BAGS IN A THERMAL ENVIRONMENT}%

\vspace*{11mm}
{\bf K.A.~Bugaev$^{1,*}$, A.I. Ivanytskyi$^1$, D.R. Oliinychenko$^{1,2}$, E.G. Nikonov$^3$, V.V.~Sagun$^1$ and G.M.~Zinovjev$^1$}\\

\vspace*{5.5mm}
$^1${Bogolyubov ITP of the National Academy of Sciences of Ukraine, Metrologichna str. 14$^B$, Kiev 03680, Ukraine}

$^2${FIAS, Goethe-University,  Ruth-Moufang Str. 1, 60438 Frankfurt upon Main, Germany}

$^3${Laboratory for Information Technologies, JINR, 141980 Dubna, Russia}\\

$^*${e-mail: bugaev@th.physik.uni-frankfurt.de}

\end{center}


\def\ts{t{\ss}s}
\def\ss{\hspace{.5pt}}
%
%
%
%

\def\i2{\mbox{\scriptsize\rm \"l\hspace*{-2.05pt}l}}

\def\bi2{\mbox{\footnotesize\rm \bf \"l\hspace*{-2.75pt}l}}

\def\ii2{\mbox{\footnotesize \it \"l\hspace*{-2.75pt}l}}

\def\I2{{\rm \"{I}}}

\newfont{\cyrfnt}{wncyr7 scaled 1400}

\newfont{\cyrB}{wncyb10 scaled 1400}

\newfont{\cyrftit}{wncyi8 scaled 985}

\newfont{\cyrfu}{wncyr8 scaled 1100}


\setcounter{page}{1}%

\noindent
{\bf Abstract.}
Here we develop the hadron resonance gas model with the Gaussian width of hadron resonances.
This model allows us to treat the usual hadrons and  the quark gluon  bags on the same footing and to study the stability of the results obtained within different formulations of the hadron resonance gas model.  In this work we  perform a successful fit of 111 independent  hadronic multiplicity ratios measured
for the center of mass collision  energies  $\sqrt{s_{NN}} =$ 2.7--200 GeV. Also we
 demonstrate that  in a narrow range of  collision energy $\sqrt{s_{NN}} =$ 4.3--4.9 GeV  there exist the peculiar   irregularities in various thermodynamic quantities found  at chemical  freeze-out.  The most remarkable irregularity  is   an unprecedented    jump of the number of effective degrees of freedom observed in this narrow energy range which is seen in all realistic versions of the hadron resonance gas model,
 including the model with 
the Breit-Wigner parameterization of the resonance width and the one  with a zero width of  all resonances.
 Therefore, the developed concept  is called the non-smooth chemical  freeze-out.  We are arguing that these irregularities evidence for the possible formation
of quark gluon  bags.  In order to develop other possible signals of their formation here we study
the apparent width of wide hadronic  resonances and quark gluon  bags in a thermal environment.
Two new effects  generated  for  the
wide resonances  and quark gluon  bags  by a thermal medium
 are discussed  here: the near threshold thermal resonance enhancement   and  the near threshold thermal  resonance sharpening. 
These effects are also analyzed for the Breit-Wigner width parameterization and it is shown that, if  the resonance decay thresholds are  located far away from the  peak  of resonance mass attenuation, then  such a width parameterization leads  to a stronger enhancement of  the resonance pressure compared to the Gaussian one. 
 On the basis of the new effects  we argue  that the most optimistic chance to find  experimentally  the quark gluon bags   may be related to their sharpening and enhancement  in  a thermal medium. In this case  the  wide quark gluon bags  may appear directly or in decays as narrow  resonances that  are absent in the tables of elementary particles  and   that have the apparent width  about 50-120 MeV  and  the  mass about or above 2.5 GeV.

\vskip 3mm

\section{Introduction}

This  year the MIT Bag model  \cite{MITBag} will celebrate its forty years anniversary.   A physical picture  about the  color degrees of freedom confined inside some  volume turned out to be
very productive and it is used as a corner stone in many subsequent phenomenological models developed in high energy nuclear physics. In particular, the   hadron bags model \cite{Kapusta:81} very efficiently exploited this idea and, compared to the statistical bootstrap model \cite{Hagedorn:65,Hagedorn:68,SBM:71},   it opened
entirely new  possibilities  to study the strongly interacting matter thermodynamics.  Since the moment of the hadron bags model formulation  \cite{Kapusta:81} its original framework was greatly extended
and now we have  a variety of exactly solvable statistical models which describe the deconfinement
phase transition  and a cross-over.  For instance, the quark gluon bag with surface tension model (QGBSTM) is able to describe
the tricritical \cite{QGBSTM1,QGBSTM1a,QGBSTM1b,QGBSTM1c} and critical endpoint \cite{QGBSTM2,QGBSTM2b} using the mechanism which is typical for ordinary
liquids.   In Ref. \cite{CGreiner:07} the authors study
an influence of an interplay between the color-flavor correlations and the definite volume fluctuations
of large quark gluon (QG) bags  on  the order of deconfinement phase transition  and the critical
endpoint properties (they consider phase transitions of higher orders), while in Ref. \cite{CGreiner:08} the previous analysis
is extended to study the role of the chiral symmetry restoration although at vanishing baryonic
density. The  work \cite{CGreiner:10} is devoted to a thorough analysis of different internal symmetries of large QG bags and to an investigation of the chiral symmetry restoration effect  on
the QCD phase diagram properties at non-vanishing baryonic densities although at the expense
of neglecting the realistic short-range repulsion between the constituents. In Ref. \cite{Ferroni:09} the novel, but rather complicated way to account for the hard-core repulsion between the QG bags is thoroughly analyzed.

 The most coherent statistical picture of quark  gluon bags is, however,   based on the finite width model  (FWM)  \cite{FWM,FWMb}. The FWM  allows one to consider these bags as heavy and wide hadronic resonances. It  takes into account not only the asymptotic spectrum of the quark gluon bags, but it also incorporates their finite and medium dependent width. The FWM  naturally explains the absence of heavy hadronic resonances in the experimental
mass spectrum compared to the Hagedorn mass spectrum \cite{Hagedorn:65}.
 Also  the FWM explains that besides the large width the QG bags are strongly  suppressed (on about fifteen to sixteen orders of magnitude compared to light hadrons!) for  temperatures below about a half of the Hagedorn temperature $T_H$, i.e for $T < \frac{1}{2} T_H$,  by the {\it subthreshold suppression}. The latter is a manifestation of  the color confinement in terms of the FWM. This property of  QG bags combined with  their   large width  and  very large number of decay channels lead to  great  difficulties in their experimental  identification (see a discussion in \cite{Blaschke:03}).

 Nevertheless, the experimental searches of  QG bags  within the existing programs and the planned ones stimulate  a  strong interest to the formulation of possible QG bag formation signals. However,
the two key questions, namely where (at what energies)  and how can  one  observe the QG bag formation, did not get  the definite answers during these almost four decades passed since
 the MIT Bag model formulation. At the same time  a discrete part of the hadronic mass spectrum of  the advanced followers \cite{QGBSTM1,QGBSTM1a,QGBSTM1b,QGBSTM1c, QGBSTM2,QGBSTM2b, CGreiner:07, CGreiner:08,CGreiner:10,Ferroni:09} of  the MIT Bag model which
is   known
as the hadron resonance gas model (HRGM)  \cite{KABCleymans:93,KABAndronic:05,KABAndronic:09,KABOliinychenko:12,KABugaev:12, KABugaev:13,Stachel:13} became
 a precise tool to extract the thermodynamic quantities at the moment of chemical freeze-out  (FO).
 The latter  is the moment at which the inelastic collisions cease to exist simultaneously for all sorts of particles.  The recent improvements of the HRGM  achieved in \cite{KABAndronic:09,KABOliinychenko:12,KABugaev:12, KABugaev:13,Stachel:13} allow one to successfully  describe all particle yield ratios measured  in the nuclear collisions at  the center of mass energies from $\sqrt{s_{NN}} = 2.7 $ GeV to  $\sqrt{s_{NN}} = 2.76 $ TeV.  Therefore, here we develop a new formulation of
 the HRGM which employes the Gaussian mass attenuation of  hadronic resonances instead of the Breit-Wigner one used in previous versions of  the HRGM.  Such a model allows us to treat the usual hadrons and  the QG bags  of  the FWM  \cite{FWM,FWMb, Reggeons:10}  on the same footing and to study the stability of the results obtained within different formulations of the HRGM. Moreover,
  a thorough analysis of  the HRGM performed here  allows us  to answer the two key questions formulated above.

  In particular, we demonstrate that  in a narrow range of  collision energy $\sqrt{s_{NN}} =$ 4.3--4.9 GeV  there exist the peculiar   irregularities in various thermodynamic quantities calculated at chemical FO.  The most remarkable irregularity  is   an unprecedented    jump of the number of effective degrees of freedom  measured in the ratios $s^{FO}/(T^{FO})^3$  (it jumps in 1.67 times)  and $p^{FO}/(T^{FO})^4$ (it jumps in 1.5 times), where $s^{FO}$,
 $T^{FO}$ and  $p^{FO}$ denote the entropy density, the temperature and the pressure taken, respectively,  at chemical FO.  In order to distinguish the present chemical FO concept from the other ones, we name it  {\it the non-smooth chemical FO}. On the basis  of the FWM,  we argue that these  irregularities, observed in all  versions of the HRGM  analyzed here, are, possibly,  related to the formation of QG bags.

 In order to answer the second key question  here we
study the modification of the wide resonances in a thermal environment and perform  a similar analysis for the QG bags. Our analysis shows  that even at  the chemical FO  a  thermal environment    essentially  modifies
the resonance mass distribution in case of large width leading  to   their  narrowing and enhancement  near  the threshold.  Based on these findings we are arguing  that the  QG  bags may be observed at the  energy range $\sqrt{s_{NN}} \simeq  $ 4.9--6 GeV  as the narrow resonances having the width of about 50-120 MeV  and  the mass about or above 2.5  which are absent in the tables of elementary particle properties.

The work is organized as follows.  The next section describes the basic equations of the suggested HRGM. In section 3 we present and discuss the results of particle yield ratios measured at the collision energies $\sqrt{s_{NN}} = 2.7-200$ GeV. The main attention is devoted to a discussion of  the found irregularities.  In section 4 the apparent width of wide resonances and QG  bags is thoroughly analyzed.
Our conclusions and some perspectives are discussed in section 5.

\section{Hadron Resonance Gas Model with Gaussian Mass Attenuation}

As a discrete part of  hadron mass-volume  spectrum the HRGM  \cite{KABCleymans:93,KABAndronic:05,KABAndronic:09,KABOliinychenko:12,KABugaev:12, KABugaev:13,Stachel:13}  is contained in all  elaborate statistical models of strongly interacting matter discussed above. In fact, it is a truncated hadronic mass spectrum of  the statistical bootstrap model \cite{Hagedorn:65,Hagedorn:68,SBM:71},  which, however, accounts for the hard-core repulsion of hadrons and their width.
The HRGM treats all  hadron resonances
known from the tables of particle properties \cite{PDG:08} with masses up to $M_0 \simeq 2.5$ GeV
as the interacting gas of Boltzmann particles.
Its basic equations  define the pressure $p  (T, \{ \mu\}) $ of such a  system with its temperature $T$ and the set of chemical potentials $\{ \mu\}$
\begin{eqnarray}
\label{EqIn}
&& p  (T, \{ \mu\})   \equiv  T \sum_k  F_k (\sigma_k) 
 \exp \left[  \frac{\mu_k - b_k p}{T} \right]   \, ,   \\
\label{EqIIn}
&&F_k (\sigma_k)   \equiv  g_k \int\limits_{0}^\infty  d m \,  \frac{\Theta\left(  m - M_k^{Th} \right) }{N_k (M_k^{Th})} 
 \exp \left[ - \frac{(m_k -m)^2}{2\, \sigma_k^2}  \right] \phi(m,  T)
 \,, \quad \\
&&\phi(m,  T)  \equiv \int \frac{d^3 p}{ (2 \pi)^3 } \exp \left[ -\frac{ \sqrt{p^2 + m^2} }{T} \right]  \, .
\label{EqIIIn}
\end{eqnarray}
In Eq. (\ref{EqIn}) the sum runs over all hadrons which include
  the following  parameters of  each   $k$-th particle: the full chemical potential of the $k$-th hadron sort  $\mu_k \equiv Q_k^B \mu_B + Q_k^S \mu_S + Q_k^{I3} \mu_{I3}$ is expressed in terms of the corresponding charges $Q_k^L$  ($Q_k^B$ is its baryonic charge, $Q_k^S$ is its strange charge and $Q_k^{I3}$ is its third isospin projection charge) and their  chemical potentials,    $m_k$ is its mean mass, $g_k$ is its degeneracy factor,  $b_k$ is its excluded  volume, while  $\sigma_k$ is the Gaussian width of this resonance which defines the true resonance width  as $\Gamma_k = Q\,  \sigma_k$ (with $Q \equiv 2 \sqrt{2\, \ln2}$) and the normalization factor is defined via the threshold mass $M_k^{Th}$   of the dominant channel as
\begin{equation}\label{EqIVn}
N_\sigma (M_k^{Th}) \equiv  \int\limits_{M_k^{Th}}^\infty  d m \,
 \, \exp \left[ - \frac{(m_k -m)^2}{2\, \sigma^2}  \right]
\end{equation}
The hard core repulsion of the Van der Waals type  generates  the suppression factor  $\exp(- b_k \,p/T )$.  $\phi (m, T) $ denotes
the thermal particle  density per  spin-isospin degree of freedom of  the  hadron sort of mass $m$.

Note that   a  new  and important  feature of the present HRGM  formulation is an inclusion of the  Gaussian width for  all hadronic resonances. Such a feature allows us to treat the usual hadrons and  the QG bags  of  the FWM which  necessarily must have the Gaussian width \cite{FWM,FWMb, Reggeons:10}  on the same footing. This is a
generalization   of    the most  successful  formulation of the HRGM
\cite{KABAndronic:05,KABAndronic:09,KABOliinychenko:12,KABugaev:12} in  which  the hadronic excluded  volumes $\{b_k\}$ are usually chosen to be equal, i.e. $b_1=b_2= ... = b_n \equiv b$.
Note that the hadron resonance gas model \cite{KABAndronic:05,KABOliinychenko:12,KABugaev:12}
with the excluded volume $ b = \frac{16}{3} \pi R^3 \simeq 0.45$ fm$^3$ and the hard-core radius $R=0.3$ fm
is able to
successfully describe  the ratios of hadronic multiplicities measured at midrapidity in the nuclear collisions
for the center of mass energies from $\sqrt{s_{NN}} = 2.7 $ GeV to  $\sqrt{s_{NN}} = 2.76 $ TeV.
The main difference of  the discrete mass-volume spectrum of the present model from the one of the most popular version of the HRGM  \cite{KABAndronic:05,KABAndronic:09,KABOliinychenko:12,KABugaev:12} is  a usage of the Gaussian width in  (\ref{EqIIIn}) instead of the
Breit-Wigner one, although the effect is claimed  to be below 10 \% even for wide hadronic resonances \cite{KABAndronic:05}.
This feature of the present model is similar to the FWM of QG bags, where
 the presence of  the Gaussian attenuation   is of a principal  importance.
 It is so because
the Breit-Wigner attenuation leads to a divergency of the  FWM partition function \cite{FWM,FWMb}.

Despite its simplicity the HRGM outlined above accurately accounts for the complexity of strong interaction between hadrons. Indeed, an attraction between them is taken into account  like in the  statistical bootstrap model  \cite{Hagedorn:65, Hagedorn:68, SBM:71} via many sorts  of hadrons  while the short range interaction  is modeled via the hard core repulsion which  leads to an appearance of  the corresponding exponentials $\exp(-  b_k \,p /T)$
 in the spectrum  (\ref{EqIn}).   In principle, a surface tension induced by an interhadron  interaction, like the one discussed recently in \cite{KABsmm:14},  should be considered for the  mass-volume spectrum of hadrons used in  (\ref{EqIn}),
but the recent  estimates made within the hadron resonance gas model \cite{KABOliinychenko:12} show that it is
small and, hence,  here it is neglected.

It is necessary to stress that the  mass attenuation in Eqs. (\ref{EqIn}) and (\ref{EqIIn}) has a clear physical meaning and it  is in line with the basic assumption of the statistical bootstrap model \cite{Hagedorn:65,Hagedorn:68,SBM:71} which suggests to account for all hadronic states with their degeneracy, that may depend on hadron mass. Therefore, the Gaussian or Breit-Wigner mass attenuations  of hadronic resonances  used in the HRGM  account for  the different hadronic states that belong to the same mass interval. This is evident, if one change the order of a hadron sort  summation and the mass integration in (\ref{EqIn})
\begin{eqnarray}
\label{EqVn}
\frac{p(T, \{ \mu\}) }{T} &\equiv & \sum_k  F_k (\sigma_k)  \exp \left[  \frac{\mu_k - b\, p}{T} \right]  \equiv\nonumber \\
& \equiv&  \int\limits_{0}^\infty  d m \,   \sum_k  \frac{\Theta\left(  m - M_k^{Th} \right) }{N_k (M_k^{Th})}
 \, \exp \left[ - \frac{(m_k -m)^2}{2\, \sigma_k^2}  \right] 
 g_k\, \phi(m,  T)  \exp \left[  \frac{\mu_k - b\, p}{T} \right] \,.
\end{eqnarray}
Before analyzing the hadronic mass spectrum in a thermal environment it is necessary to remind that the
question of  whether the experimental mass spectrum of hadrons given in the Particle Data Group tables  coincides with  the spectrum suggested by R. Hagedorn   is of great interest nowadays
\cite{KABHagedorn1,KABHagedorn2,KABHagedorn3}. However, almost all  discussions of the hadron mass spectrum simply ignore the width of  resonances, whereas  long ago it was found that
the large resonance width may essentially modify the spectrum \cite{KABStoecker:86,KABStashek:87,KABStashek:90,KABHeinz:91,KABDirk:91,Blaschke:03,FWM,FWMb, Bugaev:11jhep}. Therefore, it is important to
study the effective mass spectrum of hadrons having  a physical  width.
Another significant reason to introduce the width and decay channels of hadronic resonances into the HRGM is
that without accounting for the decays of wide resonances it is impossible to accurately describe the particle yield ratios
\cite{KABAndronic:05,KABOliinychenko:12}.
For instance, an absence of the wide $\sigma (600)$-meson which decays into two pions does not allow one to correctly describe  the pion yield since just this meson alone is responsible for about 5 \%  of pions at low AGS energy range.
Therefore, the total
particle density of hadron of sort $k$  consists of the thermal part  $n^{th}_k$  and the decay one:
\begin{eqnarray}\label{EqVIn}
n^{tot}_k & = & n^{th}_k+ n_k^{decay} = n^{th}_k + \sum_{l} n^{th}_l \, Br(l \to k) \,,\\
n^{th}_k & \equiv & \frac{\partial p}{\partial \mu_k} = \frac{F_k (\sigma_k)  \exp \left[  \frac{\mu_k - b\, p}{T} \right] }{1 + b \sum_l  F_l (\sigma_l)  \exp \left[  \frac{\mu_l - b\, p}{T} \right] } \,,
\label{EqVIIn}
\end{eqnarray}
where $Br(l \to k)$ is the decay branching ratio of  the $l$-th sort of hadron  into the hadron of sort $k$. The masses, the  widths and the strong decay branchings of all hadrons  are  taken from the particle tables  used  by  the  thermodynamic code THERMUS \cite{THERMUS}.

\begin{figure}[t]
\centerline{\includegraphics[width=88 mm, height=66 mm]{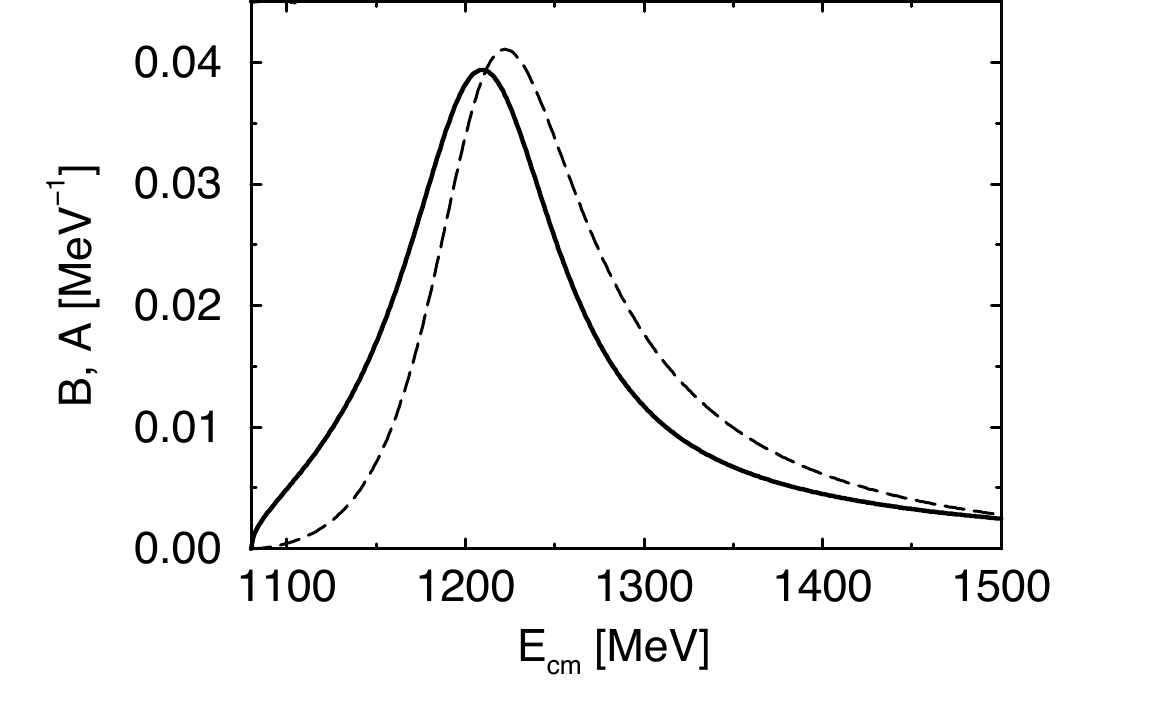}}  \vspace*{2.2mm}
\centerline{\includegraphics[width=77mm]{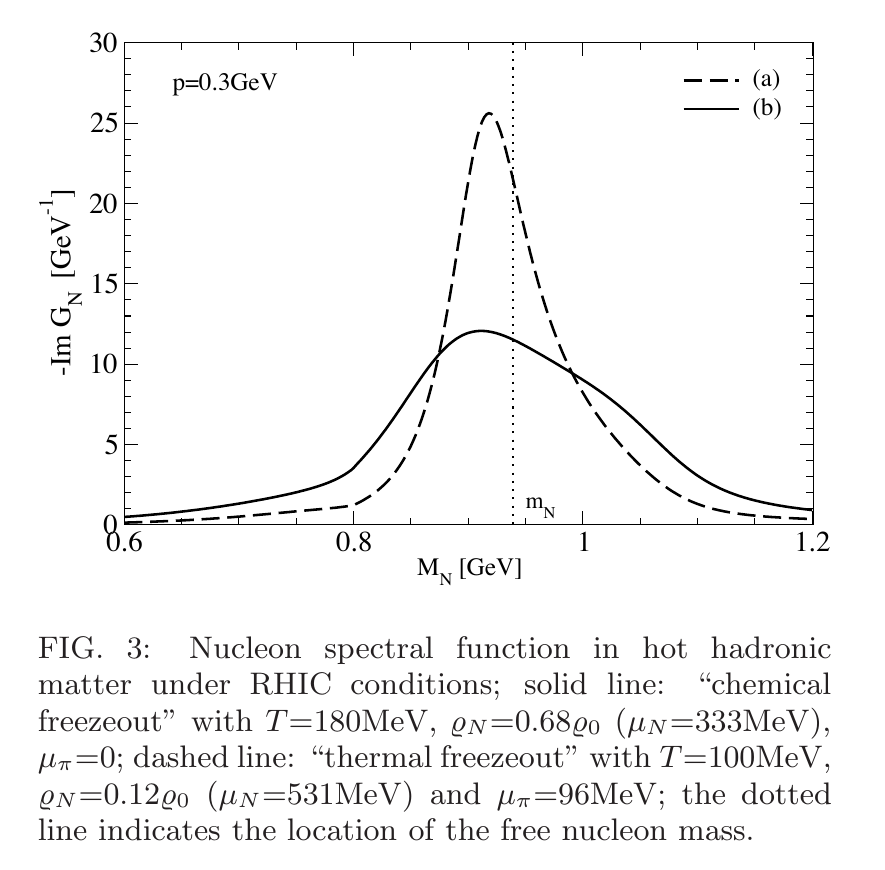}
}
 \caption{Mass dependance of the $\Delta_{33}$ resonance (upper panel) and  nucleon (lower panel) found within  effective field theoretical model at finite temperature. {\bf Upper panel:} This picture is taken from the preprint arXiv:nucl-th/9710014v2 by W. Weinhold, B. Friman and W. N\"orenberg which is published in  Ref.
  \cite{KABFriman}.
 Dashed curve denotes the mass (which is energy $E_{cm}$ in the resonance center of  mass frame) attenuation of the $\Delta_{33}$ resonance with vacuum mass and width, while the solid curve corresponds  to the low density approximation  suggested in \cite{KABFriman}.
{\bf Lower panel:} This figure is taken from the preprint arXiv:nucl-th/0407050v2 by H. van Hees and R. Rapp which is published in  Ref.  \cite{Delta:05}. As one can see from the original figure caption the authors of  Ref.  \cite{Delta:05} claim that the nucleons which are supposed to fly directly to a detector after the thermal FO, i.e. after a complete decoupling of the system, still do not have the vacuum mass and vacuum width.
 }
 \label{KABFig1}
\end{figure}

A usage of  the resonance  mass attenuation of  the Breit-Wigner (or Gaussian) type with  the  vacuum values of resonance mass and width   was heavily criticized in \cite{KABFriman}, but we find such a critique
absolutely inadequate for the states below the chemical FO.
First of  all, we note that  in the approach of \cite{KABFriman} and similar effective field theoretical models  the effect of  medium  cannot be switched off at any finite particle density or temperature.
This means that according to  the treatment of  \cite{KABFriman} and \cite{Delta:05} (and many similar works!) all the hadrons whose momentum spectra are frozen due to the absence of any strong interaction between them should keep their momentum dependent width and mass which  they acquired at the moment of kinetic FO up to they are captured by the  detectors.
Hence, according to \cite{KABFriman} all hadrons measured by detectors, including the stable ones, are some resonances that `feel' a thermal medium in which they were produced long after the medium is gone.
The typical examples of resonance mass attenuations obtained within the effective field theoretical models
are shown in Fig. \ref{KABFig1}.

This problem is well-known in the transport simulations \cite{Mosel:99,Mosel:99b}. Because  the traditional
field theoretical prescription  does not provide the correct asymptotic solutions for the particles which are stable in a vacuum it was suggested to introduce the density dependent coupling \cite{Mosel:99b} which shifts   the  particles to their  mass shell  when they propagate to the vacuum. Since there is no first principle prescription for such a procedure, we conclude that the usage of  the `crude' approximation (in terms of Ref.   \cite{KABFriman})  of  Eq.  (\ref{EqIIn})  or the Breit-Wigner one is not only possible, but it is physically adequate after the chemical FO, when
the inelastic reactions, except for the decays, cease to exist.

Second, all the `effects' which the authors of \cite{KABFriman} claim to be of principal  physical importance are reduced to a slight (by about 20 MeV) shift of the mass attenuation peak and small change of  its shape for the $\Delta_{33}$ resonance as one can see from Fig. \ref{KABFig1}.   In our opinion such modifications of the mass attenuation of   the $\Delta_{33}$ resonance compared to the `crude' approximation of  Eq.  (\ref{EqIIn})
cannot be measured  in heavy ion experiments even for such narrow a resonance as $\Delta_{33}$, since
in a vacuum  its  width is known with the accuracy of a couple of MeVs  \cite{PDG:08}.
Therefore, a serious discussion of  similar `effects'  for heavy hadronic resonances whose mass and width are  often known  with the accuracy of 100 MeV  or 200 MeV (or worse) \cite{PDG:08} does not make any sense. Thus, at the present state of  art
there is no alternative to a physically transparent Eq. (\ref{EqIIn}) to be used at and after the moment of
chemical FO. Below it will be shown, that the finite temperature affects the resulting mass distributions of resonances  much more than those tiny modification discussed in \cite{KABFriman}.
Moreover, none of  the existing field theoretical model is able to tell us what a mass attenuation
can be used for the QG bags, whereas the requirement of internal consistency of  the FWM \cite{FWM,FWMb}  leads to the Gaussian mass attenuation for the QG bags or heavy and wide resonances. Furthermore, the FWM  allows one to estimate the  parameters of the mass attenuation from the lattice QCD data \cite{FWM,FWMb} and from the experimental Regge trajectories of heavy mesons \cite{Reggeons:10}.

\section{Fit of Particle Ratios}

The fitting procedure and the choice of particle yield ratios are the same, as suggested in   \cite{KABAndronic:05,KABAndronic:09} and successfully used in  \cite{KABOliinychenko:12,KABugaev:12,KABugaev:13}. To study the effect of the Gaussian
width on the chemical FO we choose  the basic formulation of   the HRGM outlined  in \cite{KABOliinychenko:12}
for the equal  hard-core radii of all hadrons $R= 0.3$ fm which correspond to the excluded volume $b \simeq 0.45$ fm$^3$.

For the AGS  center of mass collision  energy range   $\sqrt{s_{NN}} = 2.7 -4.9$ GeV the used experimental data
 \cite{AGS_pi1, AGS_pi2} (pions),  \cite{AGS_p1, AGS_p2} (proton),   \cite{AGS_pi2} (kaons),
 \cite{AGS_L, AGS_Kas,AGS_L2,AGS_L3} (strange hyperons) and
 \cite{AGS_phi} ($\phi$ mesons)
are well-known.  As in our previous analysis  \cite{KABOliinychenko:12,KABugaev:12,KABugaev:13}
for the SPS center of mass collision  energy range   $\sqrt{s_{NN}} = 6.3 - 17$ GeV we mainly concentrate on  the NA49 Collaboration data \cite{KABNA49:17a, KABNA49:17b, KABNA49:17Ha, KABNA49:17Hb,KABNA49:17Hc,KABNA49:17phi}, which  traditionally  are the most difficult to reproduce for the HRGM. The measured  data at RHIC energies $\sqrt{s_{NN}} = 62.4, 130$ and 200 GeV are  available
from several experiments, but since the most of the data agree with each other well, we employ only the STAR data
\cite{KABstar:130a,KABstar:130b,KABstar:130c,KABstar:200a,KABstar:200b,KABstar:200c}.
More details on the fitting procedure can be found in \cite{KABOliinychenko:12,KABugaev:12,KABugaev:13}.
\begin{figure}[t]
\centerline{\includegraphics[width=88 mm]{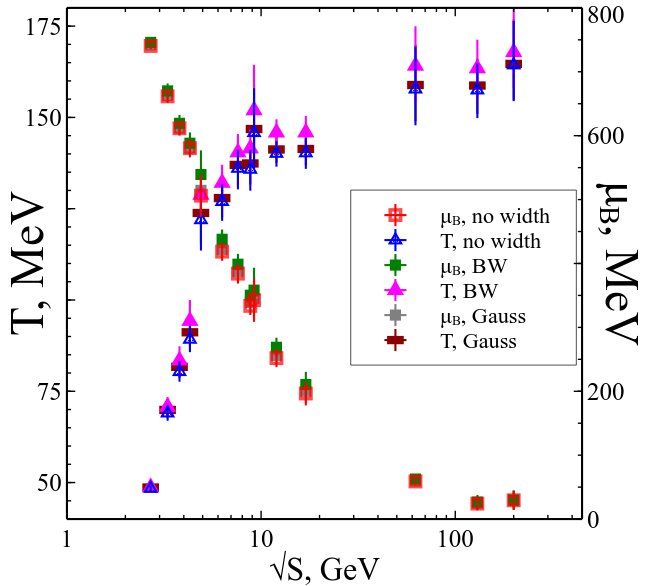}} \vspace*{2.2mm}
\centerline{\includegraphics[height=80.mm,width=88.mm]{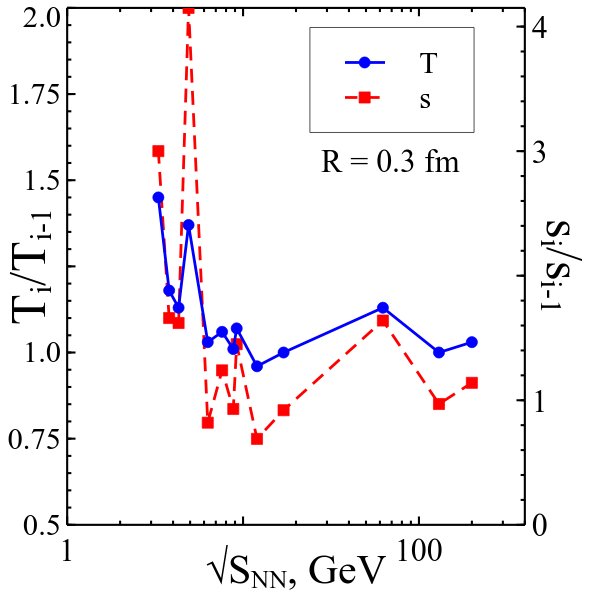}}
 \caption{{\bf Upper panel:} Center of mass energy $\sqrt{s_{NN}}$ dependence of the chemical FO temperature $T^{FO}$ and baryonic chemical potential $\mu_B^{FO}$ for
 three versions of the HRGM: with the Breit-Wigner width, with a vanishing width and with the Gaussian width.
 {\bf Lower panel:} Ratio of chemical FO temperatures $T^{FO} (\sqrt{s_{NN}(i)}/ T^{FO} (\sqrt{s_{NN}(i-1)} )$ and ratio of entropy densities $s^{FO} (\sqrt{s_{NN}(i)}/ s^{FO} (\sqrt{s_{NN}(i-1)} )$ for two subsequent energies of collision ($i\ge 2$) are shown for the HRGMG. The lines connected the symbols are given to guide the eyes.
 }
 \label{KABFig2}
\end{figure}

\begin{figure}[t]
\centerline{\includegraphics[width=7 cm]{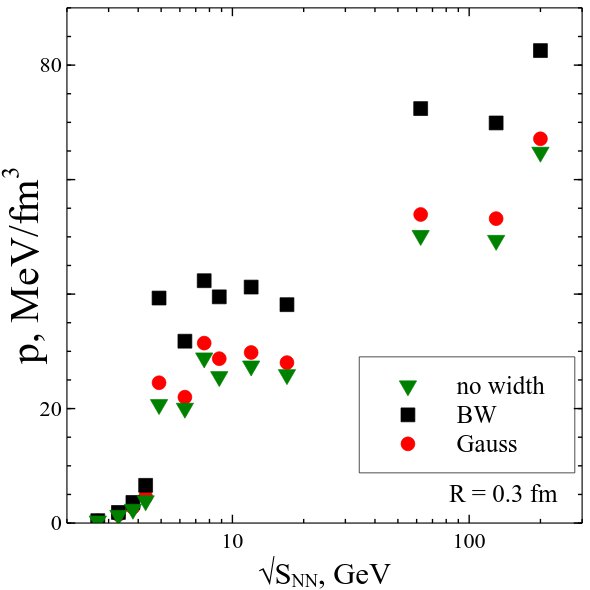}}   \vspace*{2.2mm}
\centerline{\includegraphics[width=7 cm]{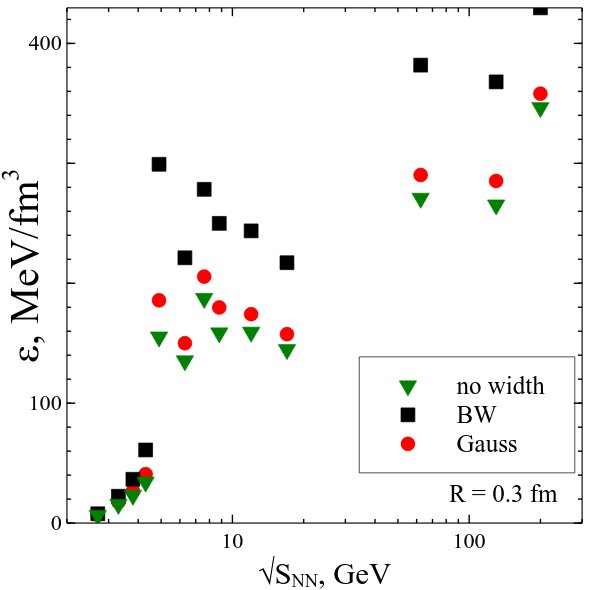}
}
 \caption{Center of mass energy $\sqrt{s_{NN}}$ dependence of the chemical FO pressure $p^{FO}$ (left) and energy density $\varepsilon^{FO}$ (right) for
 three versions of the HRGM: with the Breit-Wigner width, with a vanishing width and with the Gaussian width.
 Both quantities demonstrate a huge jump in the narrow range of collision energy $\sqrt{s_{NN}} =$ 4.3--4.9 GeV.
 The reason of why the Gaussian and zero width pressures are smaller than the Breit-Wigner one is thoroughly  analyzed in Sect. 4. }
 \label{KABFig3}
\end{figure}

As one can see from the upper  panel of Fig. \ref{KABFig2} the chemical FO temperature $T^{FO}$ and baryonic chemical potential
$\mu_B^{FO}$  of the present model,
which hereafter is called HRGMG, for convenience,
 are more close to the  HRGM with a zero width (HRGM0) than to the HRGM with the Breit-Wigner mass attenuation (HRGMBW).
Moreover, the deviation of the chemical FO temperatures obtained  within HRGMG  and within HRGMBW slowly grows  with the increase of  collision energy, but their  difference does not 6 MeV.
From the lower panel of  Fig. \ref{KABFig2} it is clearly seen that in the narrow range of collision energy $\sqrt{s_{NN}} =$ 4.3--4.9 GeV the chemical FO temperature $T^{FO}$ increases in about 1.35 times,
while the entropy density at chemical FO in this case jumps in about 4.2 times!
A similar picture is seen in Fig. \ref{KABFig3}  for the  pressure $p^{FO}$ and energy density $\varepsilon^{FO}$ at chemical FO.  We would like to note that these quantities $s^{FO}$, $p^{FO}$ and $\varepsilon^{FO}$ demonstrate a remarkable irregularity in the narrow range of collision energy $\sqrt{s_{NN}} =$ 4.3--4.9 GeV.
Thus, from Figs. \ref{KABFig2} and \ref{KABFig3} one can see that for the HRGMG  the chemical FO pressure increases in 5 times,
the energy density jumps in about 4.75 times while the chemical FO temperatures changes in about 1.35 times, when the
collision energy increases from $\sqrt{s_{NN}} =$ 4.3 GeV to $\sqrt{s_{NN}} =$ 4.9 GeV, i.e. it grows on about 14 \% only. In other words, the quantities  $s^{FO}/(T^{FO})^3$  and $p^{FO}/(T^{FO})^4$ which usually characterize  the number of effective degrees of freedom increase, respectively,  in about 1.67 and 1.5 times, while the collision energy  changes on  about 14 \%.
For the HRGMBW  and for the unrealistic  HRGM0 the results are very similar, although 
the  pressure  and the  energy density  obtained by   the HRGMBW for the corresponding values of the collision energy  are essentially larger 
than the ones  found by  the HRGMG. In Section 4  it is shown that the main reason for stronger 
HRGMBW pressure compared to the HRGMG one is due to the width parameterization. 

Note that a similar behavior of  $T^{FO}$ and a large decrease of the chemical FO volume at this energy range was found in  \cite{KABAndronic:05} for the HRGM with a vanishing width, however the authors of  \cite{KABAndronic:05} never discussed  the strong irregularities in the  chemical FO pressure and energy density and their reduced values $p^{FO}/(T^{FO})^4$ and  $\varepsilon^{FO}/(T^{FO})^4$.

\begin{figure}[t]
\centerline{\includegraphics[width=70mm,height=7cm]{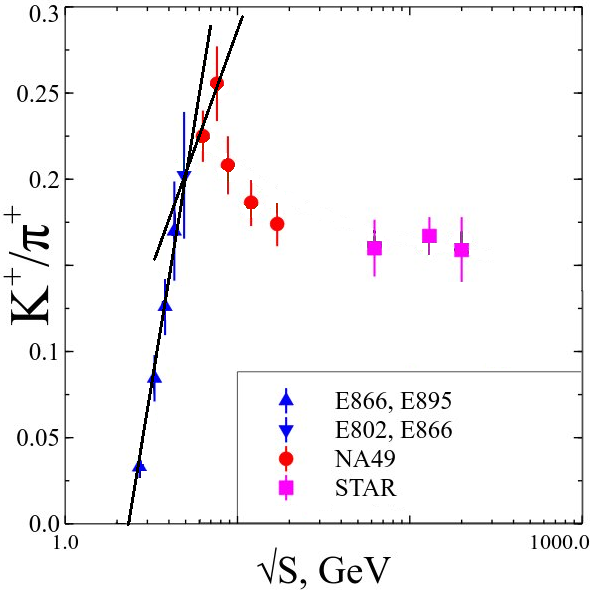}}  \vspace*{2.2mm}
\centerline{\hspace*{8.8mm}\includegraphics[width=77mm,height=7cm]{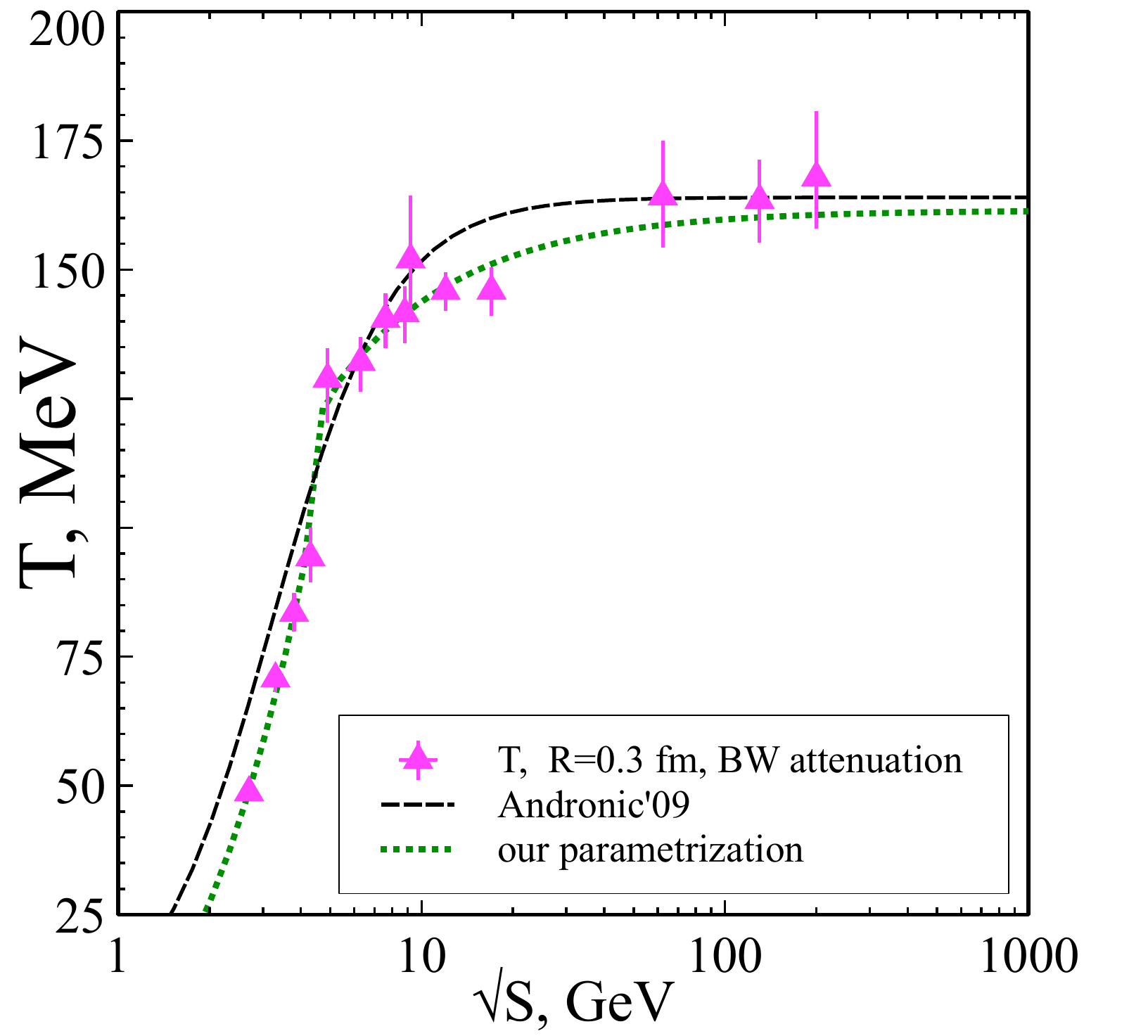}
}
 \caption{\footnotesize {\bf Upper panel:} The center of mass energy $\sqrt{s_{NN}}$  dependence of the experimental  $K^+/\pi^+$ ratio (Strangeness Horn). The straight lines indicate the change of slope from 0.667 for  $\sqrt{s_{NN}}\in [2.7; 4.9] $ GeV to 0.275  for  $\sqrt{s_{NN}}\in [4.3; 7.6] $ GeV.
 {\bf Lower panel:} The  chemical FO temperature  $T^{FO}$ for the HRGMBW as a function of the center of mass energy $\sqrt{s_{NN}}$ and its two theoretical representations. The dashed curve corresponds to Eq. (\ref{EqXN}),
while the dotted curve represents Eq. (\ref{EqXIIN}).
 }
 \label{KABFigX}
\end{figure}

It is interesting that the  irregularities found here   are also accompanied by the irregularity in the  $\sqrt{s_{NN}} $ dependence of the Strangeness Horn \cite{Horn}, i.e. the multiplicity  ratio  $K^+/\pi^+$.
As one can see from the straight lines in the upper  panel of Fig.  \ref{KABFigX}, there is a strong  change of  the $K^+/\pi^+$ ratio slope at the energy range  $\sqrt{s_{NN}} \simeq 4.3-4.9$ GeV.  Of course, the  existing  large error bars do not allow us to locate the transition point with the high accuracy, but it is hoped that the future experiments will provide us with essentially smaller errors which, in turn, will allow one to make more definite conclusions.
 It is odd, that from the very beginning all found dependences
of thermodynamic quantities at  chemical FO were assumed to be continuous and  smooth  \cite{Cleymans:06,KABAndronic:05}.  Probably, such an attitude did not allow other researchers to find out these irregularities.
In contrast to the traditional  assumptions   Figs. \ref{KABFig1}, \ref{KABFig2} and \ref{KABFig3} clearly demonstrate
that the  $\sqrt{s_{NN}} $ dependence of  chemical FO parameters $T^{FO}$, $p^{FO}$ and  $\varepsilon^{FO}$ has a discontinuity.
Therefore, in order to distinguish our concept of chemical FO from the previous findings  with a smooth functional dependences we name it
as {\it the non-smooth chemical FO}.

In order to parameterize the
  functions  $T^{FO} (\sqrt{s_{NN}})$ and   $T^{FO} (\mu_B^{FO})$  for the non-smooth chemical FO,
let us,  for convenience, introduce the functions
\begin{eqnarray}\label{EqVIIIN}
c_+(x,a,b) &=& \frac{1}{1+e^{(x-a)/b}} = \frac{1}{2}\left(1+\tanh{\frac{x-a}{2b}}\right)\\
c_-(x,a,b) &=& \frac{1}{1+e^{(a-x)/b}} = \frac{1}{2}\left(1-\tanh{\frac{x-a}{2b}}\right)
\label{EqIXN}
\end{eqnarray}
Such functions, sigmoids,   are well known in physics because they represent Bose-Einstein distributions. In the limit of small values of the parameter $b$ they become  $c_+(x,a,b)|_{b \to 0} = \theta(a-x)$  and
 $c_-(x,a,b)|_{b \to 0} = \theta(x-a)$, where $\theta(x)$ is the usual Heaviside function. For finite $b$ values the functions (\ref{EqVIIIN}) and
(\ref{EqIXN}) cut the $x$ values for  $x>a$  and  $x<a$, respectively.
The  width of a smooth cut transition is of about $2b$.

\begin{figure}[t]
\centerline{\includegraphics[width=7.7 cm,height=7cm]{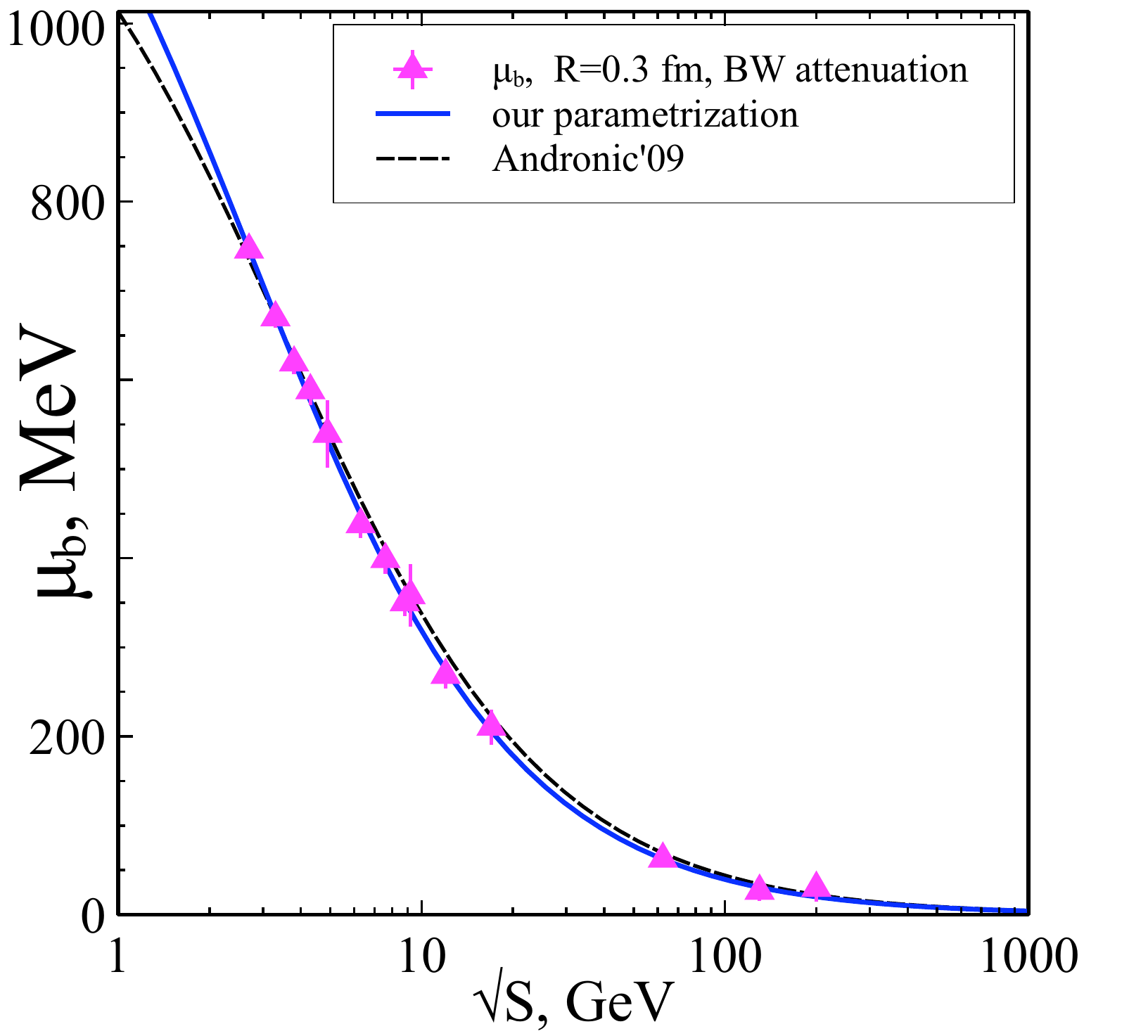}}   \vspace*{2.2mm}
\centerline{\includegraphics[width=7.7 cm,height=7cm]{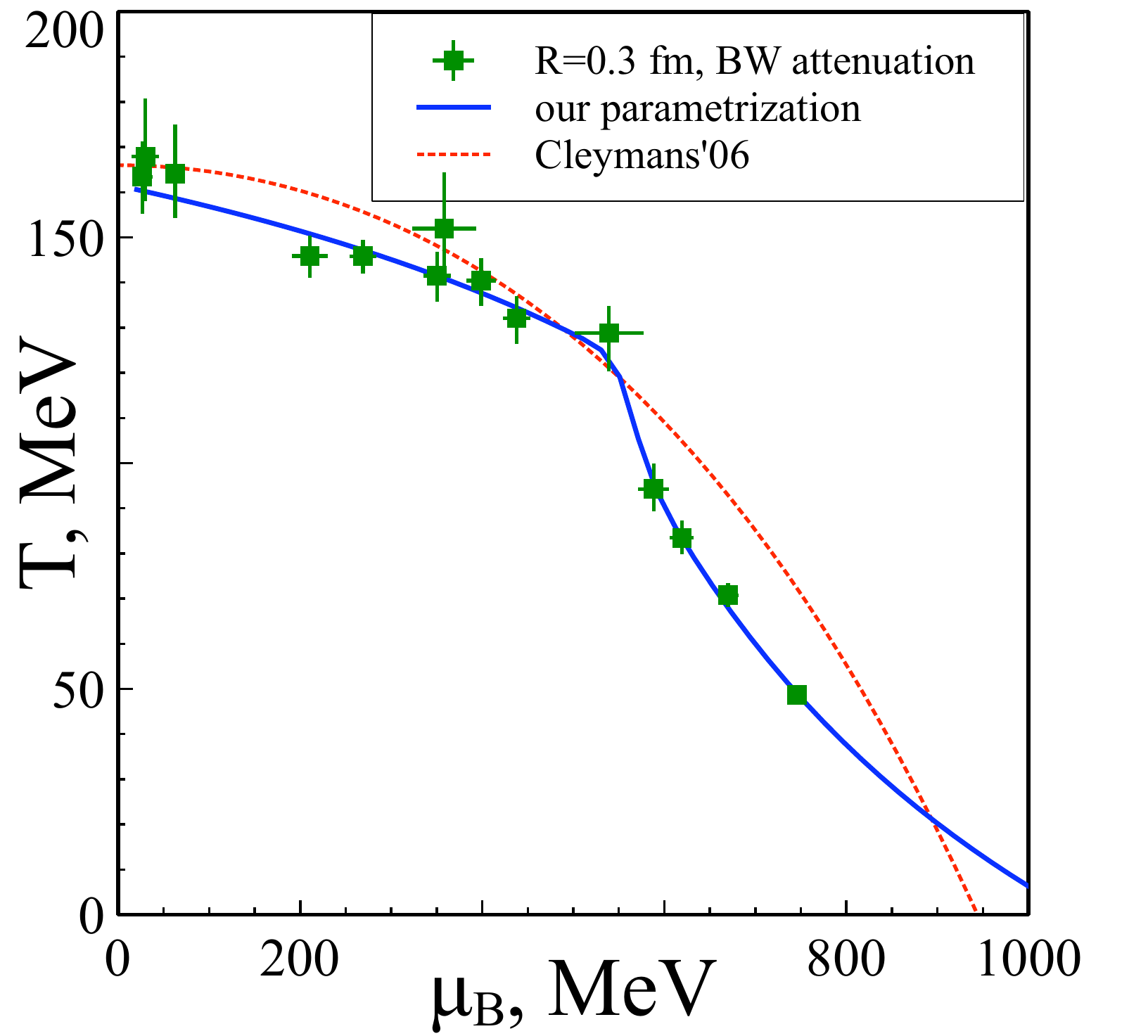}
}
 \caption{\footnotesize  {\bf Upper panel:} The baryonic chemical potential   $\mu_B^{FO}$ at chemical FO   as a function of the center of mass energy $\sqrt{s_{NN}}$ for the HRGMBW. The dashed  and   solid    curves  are, respectively,  defined by Eqs.  (\ref{EqXIN})  and  (\ref{EqXIIIN}).
 {\bf Lower panel:} The chemical FO temperature  $T^{FO}$
 as a function of the baryonic chemical potential at FO  $\mu_B^{FO}$
 for the HRGMBW.  The solid  and dotted   curves  are, respectively,  defined by  the system   (\ref{EqXIIN}), (\ref{EqXIIIN})  and  by Eq.  (\ref{EqXIVN}).
}
 \label{KABFig4N}
\end{figure}

First let us discuss some popular parameterizations   for $T^{FO} (\sqrt{s_{NN}})$ and   $\mu_B^{FO}(\sqrt{s_{NN}})$.
In 2009 Andronic and collaborators in \cite{KABAndronic:09}
considered slightly different particle spectrum in their HRGM  developed in \cite{KABAndronic:05}  (in particular introduced $\sigma$ - meson )
in order to improve their results of  Ref. \cite{KABAndronic:05} on pion multiplicity and, hence, on the Strangeness Horn.
Their thermal parameters have altered, but only very slightly. In \cite{KABAndronic:09} they suggested the following parameterizations for  $T^{FO} (\sqrt{s_{NN}})$ and   $\mu_B^{FO}(\sqrt{s_{NN}})$
\begin{eqnarray}\label{EqXN}
T^{FO}[MeV]  &= & \frac{T^{lim}_A}{1 + \exp \left[  2.60 - \ln(\sqrt{s_{NN}})/0.45    \right] } =  T^{lim}_A \, c_-(\ln(\sqrt{s_{NN}}), 1.17, 0.45) \,, \\
\mu_B^{FO}[MeV] &= &  \frac{a_A}{1 + b_A \sqrt{s_{NN}} }\, ,
\label{EqXIN}
\end{eqnarray}
with $T^{lim}_A = 164\pm 4$ MeV,  $a_A = 1303 \pm 120 $\, MeV, $b_A = 0.286 \pm 0.049 $\, GeV$^{-1}$
and $\sqrt{s_{NN}}$ is given in GeVs.
The  functions  $T^{FO} (\sqrt{s_{NN}})$  (\ref{EqXN}) and $\mu_B^{FO}(\sqrt{s_{NN}})$
(\ref{EqXIN}) are, respectively, shown by the dashed curves in the lower  panel of Fig. \ref{KABFigX} and in the upper  panel of Fig. \ref{KABFig4N}. As one can see from the upper    panel of Fig. \ref{KABFig4N}
the function $\mu_B^{FO}(\sqrt{s_{NN}})$ (\ref{EqXIN}) rather well describes the chemical FO points of the HRGMBW,
but the function  $T^{FO} (\sqrt{s_{NN}})$  (\ref{EqXN}) describes well only the points with the collision energy $\sqrt{s_{NN}} = 9.2, 62.4, 130$ and 200 GeV, while the other points, especially for $\sqrt{s_{NN}} \le  4.9$ GeV,  are poorly reproduced.

We suggest another type of parametrization, which  aims at a precise description of the chemical FO points. We notice that the  behaviour of $T^{FO}(\sqrt{s_{NN}})$ is qualitatively different in the regions  $\sqrt{S} < 4.5$ GeV and  $\sqrt{S} > 4.5$ GeV.  Therefore, the dependences  $T^{FO}(\sqrt{s_{NN}})$  and
$\mu_B^{FO}(\sqrt{s_{NN}})$ found by the HRGMBW are parameterized  in the following way:
\begin{eqnarray}\label{EqXIIN}
&&T^{FO}[MeV] = (T_{1O}+T_{2O}\sqrt{s_{NN}}) c_+(\sqrt{s_{NN}},4.5,0.1)  +  (T_{3O}/\sqrt{s_{NN}} + T^{lim}_O) c_-(\sqrt{s_{NN}},4.5,0.1) \, , \\
&&\mu_B^{FO}[MeV] = \frac{a_O}{(1 + b_O \sqrt{s_{NN}})} \,.
\label{EqXIIIN}
\end{eqnarray}
Here the functions $c_+$ and $c_-$ are used  to make a smooth transition from one kind of  the  $\sqrt{s_{NN}}$ behavior to another. The fitting results in the following values of the coefficients
$T_{1O} = -34.4$ MeV, $T_{2O} = 30.9$ MeV/GeV,
$T_{3O} = -176.8 $ GeV$\cdot$MeV, and
$T^{lim}_O  = 161.5$  MeV
with   $\chi^2/dof = 6.6/10$ for  $T^{FO} (\sqrt{s_{NN}})$ (\ref{EqXIIN}). The parameterization for $\mu_B^{FO}(\sqrt{s_{NN}})$ (\ref{EqXIIIN})  is the same as in (\ref{EqXIN}), but the coefficients are different, i.e.
$a_O = 1481.6 $\, MeV, $b_O = 0.365 $\, GeV$^{-1}$.
For the parametrization  (\ref{EqXIIIN}) we found  $\chi^2/dof$ = 2.3/12. The resulting curve $T^{FO} (\mu_B^{FO})$ (see the solid curve in the lower  panel  of Fig. \ref{KABFig4N}) has
$\chi^2$ = 4.9 for 14 data points  and it does not contain any free parameter.
Note that for the chemical FO curves of the HRGMG the parameters entering Eqs. 
(\ref{EqXIIN})  and  (\ref{EqXIIIN}) are practically the same and, hence, we do not show these curves    in the figures. 
 From the upper panel of Fig. \ref{KABFig2} it is easy to understand 
the fact that the main difference between the parameters of  curves  (\ref{EqXIIN})  and  (\ref{EqXIIIN}) corresponding to the  HRGMBW and the HRGMG is in the value of $T^{lim}_O$,
which is about  6 MeV smaller for the latter model.
We hope that the parameterizations   (\ref{EqXIIN})  and  (\ref{EqXIIIN}) can be verified with the more precise  data which  will be measured in a few years on new accelerators, since the narrow range in the center of mass energy of collision corresponds to a wide range of the laboratory energy.

In the lower  panel of Fig.  \ref{KABFig4N} we compare the resulting curve $T^{FO} (\mu_B^{FO})$ (solid curve) obtained in this work with the parameterization
\begin{eqnarray}\label{EqXIVN}
T[GeV] = 0.166 - 0.139 \left(\mu^{FO}_B \right)^2 - 0.053\left( \mu^{FO}_B\right)^4 \,,
\end{eqnarray}
suggested in \cite{Cleymans:06}.  Eq. (\ref{EqXIVN}) is a fit to  the chemical FO points obtained by different versions of the HRGM. Several of these HRGM analyzed in \cite{Cleymans:06} take into account the hard core repulsion,  but none of them included  all hadronic states and none of them considered   the  width of hadronic resonances in the full range of collision energy. Now it is clear  that these two approximations cause the large deviation  from  the results obtained in this work.

From the lower  panel of Fig.  \ref{KABFig4N} one can see that at small baryonic chemical potentials  the slopes of the curves   defined by Eq. (\ref{EqXIVN}) and by a system (\ref{EqXIIN}), (\ref{EqXIIIN}) are different. Note that there is no a priori reason to believe that the function $T^{FO} (\mu_B^{FO})$ 
should have a vanishing $\mu_B^{FO}$ derivative  at $\mu_B^{FO} = 0$. The high quality description 
of the chemical FO points provided by a system (\ref{EqXIIN}), (\ref{EqXIIIN})  gives an evidence against such a belief. 

It is interesting to note that the traditional  functions   $T^{FO} (\mu_B^{FO})$ like the one given by Eq. (\ref{EqXIVN})  which describe the smooth dependence of  chemical FO parameters  are often used in the works employing the obsolete and oversimplified versions of the HRGM. The latter versions have very short  list of hadronic states and they do not account for the resonance decays and
for their nonzero width. In fact, the vast majority of such models was  never used to describe the actual experimental data, but they are employed for various `predictions'. Therefore, the likelihood of such  `predictions' is very low. Typical examples  of such works are  Refs. \cite{Begun:13, Begun:13b}, where the authors in detail  analyze the ratio of the entropy density to the cube of temperature at chemical FO, i.e.   $s^{FO}/(T^{FO})^3$,  using a smooth parameterization $T^{FO} (\mu_B^{FO})$ (\ref{EqXIVN}) (see, for instance,  Fig. 4b in \cite{Begun:13}). As one can see from Fig. 4b in \cite{Begun:13} the sum of  ratios $s^{FO}/(T^{FO})^3$ found for mesons and baryons should demonstrate strong decrease  as a function of  collision energy  at $E_{lab} \ge 3$ GeV, i.e at  $\sqrt{s_{NN}} \ge $ 2.7 GeV.
 Note that such a behavior  is not seen within the realistic versions of the HRGM.

\begin{figure}[t]
\centerline{\includegraphics[width=77 mm]{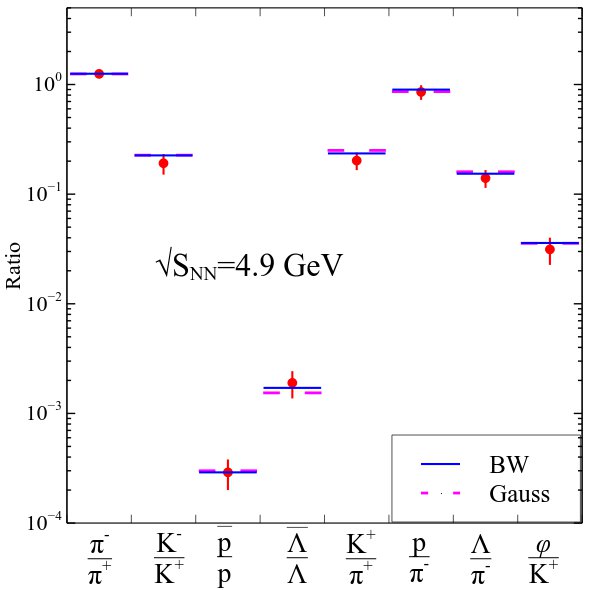}} \vspace*{2.2mm}
\centerline{\includegraphics[width=77.mm]{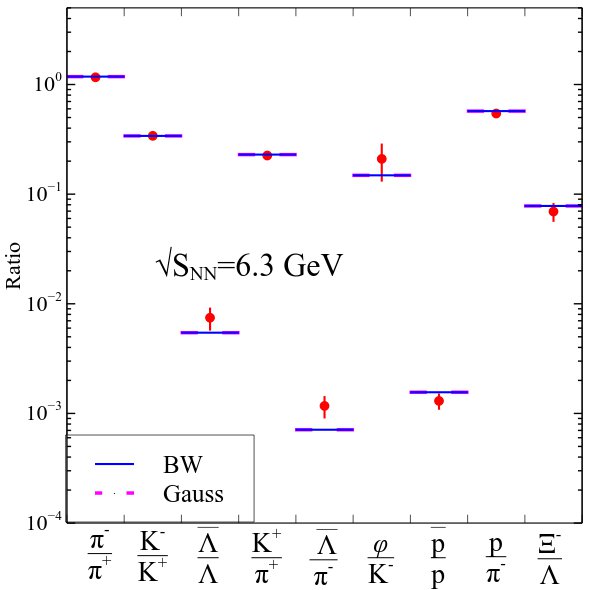}}
 \caption{The particle  yield  ratios described   by  the present HRGMG. The best fit for   $\sqrt{s_{NN}} = 4.9$ GeV is obtained  for   $T \simeq 123.8$ MeV, $\mu_B \simeq 514.4$ MeV,  $\mu_{I3} \simeq -16.75$ MeV (upper panel), whereas  for  $\sqrt{s_{NN}} = 6.3$ GeV (lower panel) it is obtained for
 $T \simeq 127.9$ MeV, $\mu_B \simeq 421.9$ MeV,  $\mu_{I3} \simeq -10.9$ MeV.  A yield ratio of two particles is denoted by the ratio of their respective symbols.
}
 \label{KABFig4}
\end{figure}
\begin{figure}[t]
\centerline{\includegraphics[width=77 mm]{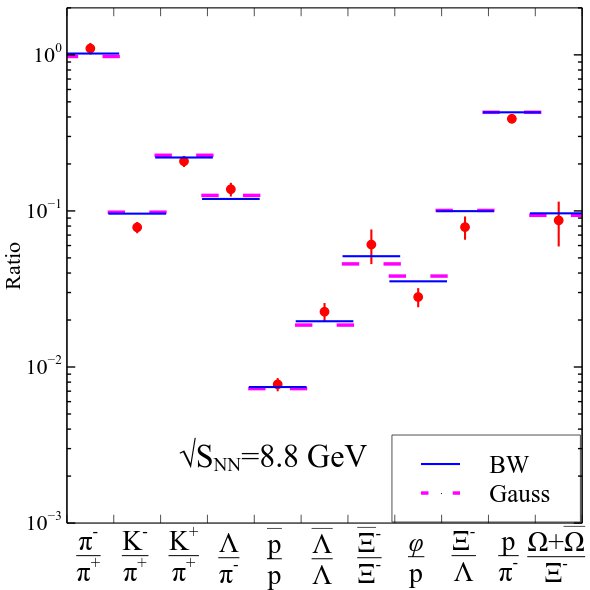}} \vspace*{2.2mm}
\centerline{\includegraphics[width=77.mm]{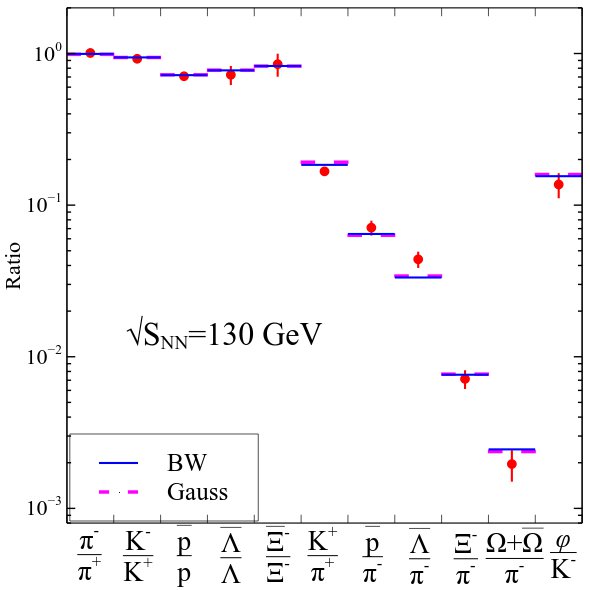}}
 \caption{Same as in Fig.  \ref{KABFig4}, but for $\sqrt{s_{NN}}= $ 8.8 and 130 GeV.
 For  $\sqrt{s_{NN}}= $ 8.8 GeV the fit parameters are $T \simeq 137.4$ MeV, $\mu_B \simeq 338.5$ MeV,  $\mu_{I3} \simeq 2.19$ MeV (upper panel), whereas  for  $\sqrt{s_{NN}} = 130$ GeV (lower panel) one finds
 $T \simeq 158.7$ MeV, $\mu_B \simeq 24.93$ MeV,  $\mu_{I3} \simeq 1.64$ MeV.
  }
 \label{KABFig5}
\end{figure}

A few  selected particle yield ratios  described within the HRGMG are shown in Figs. \ref{KABFig4}
and \ref{KABFig5}. From these figures one can see that overall fit is good and that the
results of   the HRGMG are almost the same as the ones found  within the HRGMBW.  The resulting $\chi^2/dof$ obtained for the HRGMG is  $\chi^2/dof = 150.8/59 \simeq 2.56$. It is only 17 \% larger than
the one $\chi^2/dof = 125.5/59 \simeq 2.12$ found for  the  HRGMBW.  From our previous experience  we know that the main part of these values of  $\chi^2/dof$  should be attributed
to the ratios involving  the heavy strange particles \cite{KABOliinychenko:12,KABugaev:12, KABugaev:13}
(see, for instance, the ratios $\Lambda/\pi^-$, $\bar \Lambda/\pi^-$ and $\phi/p$ in Figs. \ref{KABFig4} and \ref{KABFig5}). However, in Refs.  \cite{KABOliinychenko:12,KABugaev:12, KABugaev:13} it was also demonstrated that the quality of the fit  of these problematic ratios can be essentially improved, if one uses the multi-component hard-core repulsion \cite{KABugaev:12} and introduces the strangeness enhancement factor  \cite{KABugaev:13}. An additional important
feature of such improvements is that they practically do not affect  the values of thermodynamic
functions at chemical FO since the variation of  $T^{FO}$ and $\mu_B^{FO}$ is small and
the contribution  of heavy strange particles into  $s^{FO}$, $p^{FO}$ and $\varepsilon^{FO}$ is almost negligible.  Moreover, a close inspection of  the thermodynamic quantities
$s^{FO}$, $p^{FO}$ and $\varepsilon^{FO}$ obtained within the most successful versions of
the HRGM \cite{KABOliinychenko:12,KABugaev:12, KABugaev:13} shows that the non-smooth
chemical FO  is  their inherent feature.

Now we would like to  clarify the question  whether the found irregularities and a drastic change in the number of effective degrees of freedom  seen at $\sqrt{s_{NN}} =$ 4.3--4.9 GeV  are   related to the deconfinement transition from hadron gas to QG plasma.
Although  the famous irregularities known as the Kink \cite{Kink}, the Strangeness Horn
\cite{Horn} and the Step \cite{Step}  are observed at somewhat higher energy $\sqrt{s_{NN}} \ge $ 7.63 GeV,
we have to  point out   that before  more convincing signals of the deconfinement phase transition   will be found   there is no reason to believe that the mentioned  irregularities \cite{Kink,Horn,Step}  are, indeed,  the signals of the onset of deconfinement. Moreover,  despite the widely spread  claims \cite{Claims}
an absence of solid theoretical backup  of these irregularities \cite{Kink,Horn,Step} does not allow
one to consider them as  the convincing signals of the onset of deconfinement.
Furthermore, a successful description of every tiny detail  of the  Strangeness Horn achieved
recently within the  HRGMBW  \cite{KABugaev:13} tells us that there are no associated  irregularities
in the thermodynamic quantities at and above  the corresponding value of the collision energy $\sqrt{s_{NN}} = $ 7.63 GeV.  Therefore, before more convincing signals are found  the irregularities at $\sqrt{s_{NN}} =$ 4.3--4.9 GeV reported above
 can be also considered as the signals of the deconfinement phase transition.

In  contrast to some general speculations \cite{Claims}  on a possible source  of  the above mentioned  irregularities \cite{Kink,Horn,Step}, the non-smooth chemical FO has additional theoretical backup provided by the  FWM \cite{FWM, FWMb,Reggeons:10}.  The FWM \cite{FWM,FWMb} is able to successfully describe a variety of the lattice QCD thermodynamics data \cite{LQCD:1,LQCD:2,LQCD:3}
at vanishing baryonic chemical potential. Also its predictions for the Regge trajectories of non-strange and strange mesons \cite{Reggeons:10}  were  successfully confirmed  by the
thorough  analysis of  both the real and the imaginary parts of
the leading Regge trajectories  \cite{Bugaev:11jhep} of
$\rho_{J^{--}}$, $\omega_{J^{--}}$, $a_{J^{++}}$ and  $f_{J^{++}}$  mesons for the spin values  $J \le 6$
and the ones of  $K^*_{J^{P}}$ mesons of isospin $\frac{1}{2}$,  parity $P=(-1)^J$  and  spin values  $J \le 5$. One of the most important predictions  of the  FWM \cite{FWM,FWMb} is  that at vanishing baryonic chemical potential   the  QG bags are  strongly suppressed (by a factor of fifteen-sixteen order of magnitude)
compared to nucleons up to the temperatures of  half of Hagedorn temperature, i.e. $T_{suppr} \simeq \frac{1}{2} T_H$.  Since for the different lattice QCD data analyzed in \cite{FWMb} the  Hagedorn temperature value varies from $T_H \simeq 160$ MeV to
$T_H \simeq 188$ MeV, then the corresponding range of the suppression temperatures is $T_{suppr} \simeq 80 - 95$ MeV. It is remarkable that the chemical FO temperatures at  $\sqrt{s_{NN}} =$ 4.3  GeV
are $T_{FO} \simeq 90-95$ MeV, while at $\sqrt{s_{NN}} =$ 4.9  GeV they are
$T_{FO} \simeq 123-128$ MeV. In other words, according to the FWM
\cite{FWM,FWMb} at $\sqrt{s_{NN}} =$ 4.9  GeV the QG bags can be  formed.
Of course, at the moment it is unclear whether the QG bags are formed at the chemical FO stage as the metastable states of finite systems created in nuclear collisions or  they are formed at earlier stages of collision.

Also one may dispute our argument to apply to high baryonic densities  the results obtained within the FWM at the vanishing ones. We, however, should point out  that the density of states of QG bags with zero baryonic charge, i.e. meson-like bags, cannot depend on the baryonic charge of  the system.
The question  about  the baryonic density dependence  of the mass density and the width of meson-like  QG bags cannot be directly answered at the moment. Nevertheless, one should  account  for  two facts. The first of them  is   that  the FWM establishes a strict  proportionality between the pseudo-critical temperature $T_{pc}$ at the cross-over and the Hagedorn
temperature $T_H = c \,T_{pc}$  with the constant $c \in [0.92-0.98]$ which weakly  depends on the number of quark flavors  and number of  colors.   The second fact is  that the lattice QCD thermodynamics shows
 very weak dependence of   $T_{pc} $ on the baryonic chemical potential $\mu_B$ \cite{BNL-B}:
$$
\frac{T_{pc}(\mu_B)}{T_{pc}(0)} = 1 - 0.0066(7) \left(  \frac{\mu_B}{T_{pc}(\mu_B)} \right)^2\,.
$$
Therefore, from these facts we conclude that  the possible  $\mu_B$ dependences  of the mass density and the width of  meson-like QG bags should be very weak as well. Thus, our previous estimates on the Hagedorn temperature value of meson-like QG bags  should be valid at nonzero $\mu_B$ values as well.

In the next section we discuss the question how can one possibly observe the meson-like QG bags.

\section{An apparent width of wide resonances and meson-like QG bags}

Now we would like to analyze the apparent width of  wide resonances and meson-like QG bags in a thermal environment.
 The typical term of  the $k$-resonance that enters into  the mass spectrum (\ref{EqIn})  of the HRGMG  is as follows
\begin{eqnarray}
\label{EqVIIIn}
%
&&\hspace*{-0mm}F_k (\sigma_k)  \equiv   g_k \int\limits_{0}^\infty  d m \,  \frac{\Theta\left(  m - M_k^{Th} \right) }{N_k (M_k^{Th})}
 \exp \left[ - \frac{(m_k -m)^2}{2\, \sigma_k^2}  \right]  \int \frac{d^3 p}{ (2 \pi)^3 }   \exp \left[ -\frac{ \sqrt{p^2 + m^2} }{T} \right] . \quad \quad 
\end{eqnarray}
%
The notations used  in (\ref{EqVIIIn}) are the same as in Eqs. (\ref{EqIIn})--
(\ref{EqIVn}). Evidently, 
for the narrow resonances the term $F_k (\sigma_k)$ converts into the usual thermal density of particles, i.e. for $\sigma_k \rightarrow 0$ one has $F_k \rightarrow  g_k \, \phi(m_k, T)$.

\begin{figure}[t]
\centerline{\includegraphics[width=7 cm]{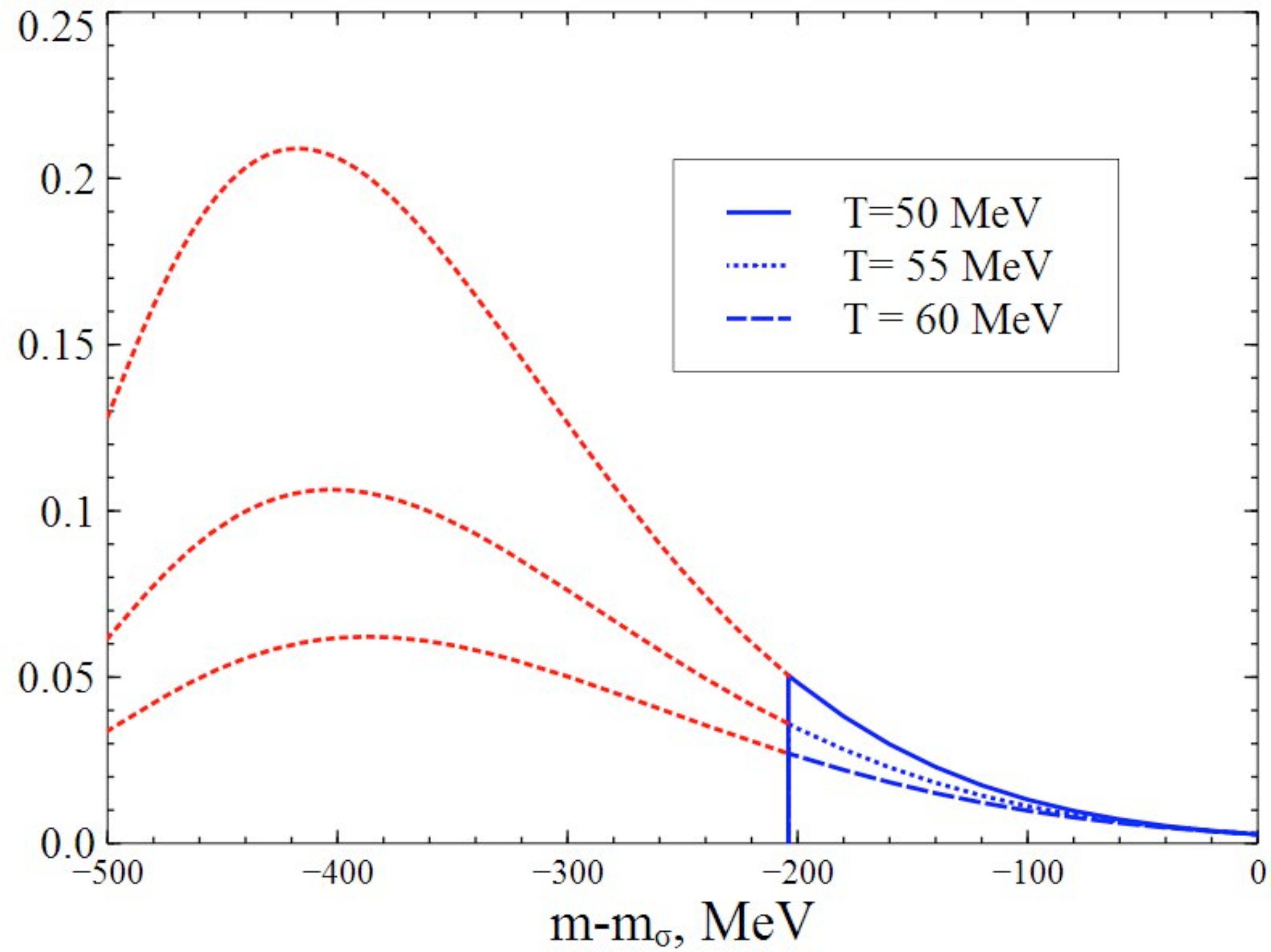}}   \vspace*{2.2mm}
\centerline{\includegraphics[width=7 cm]{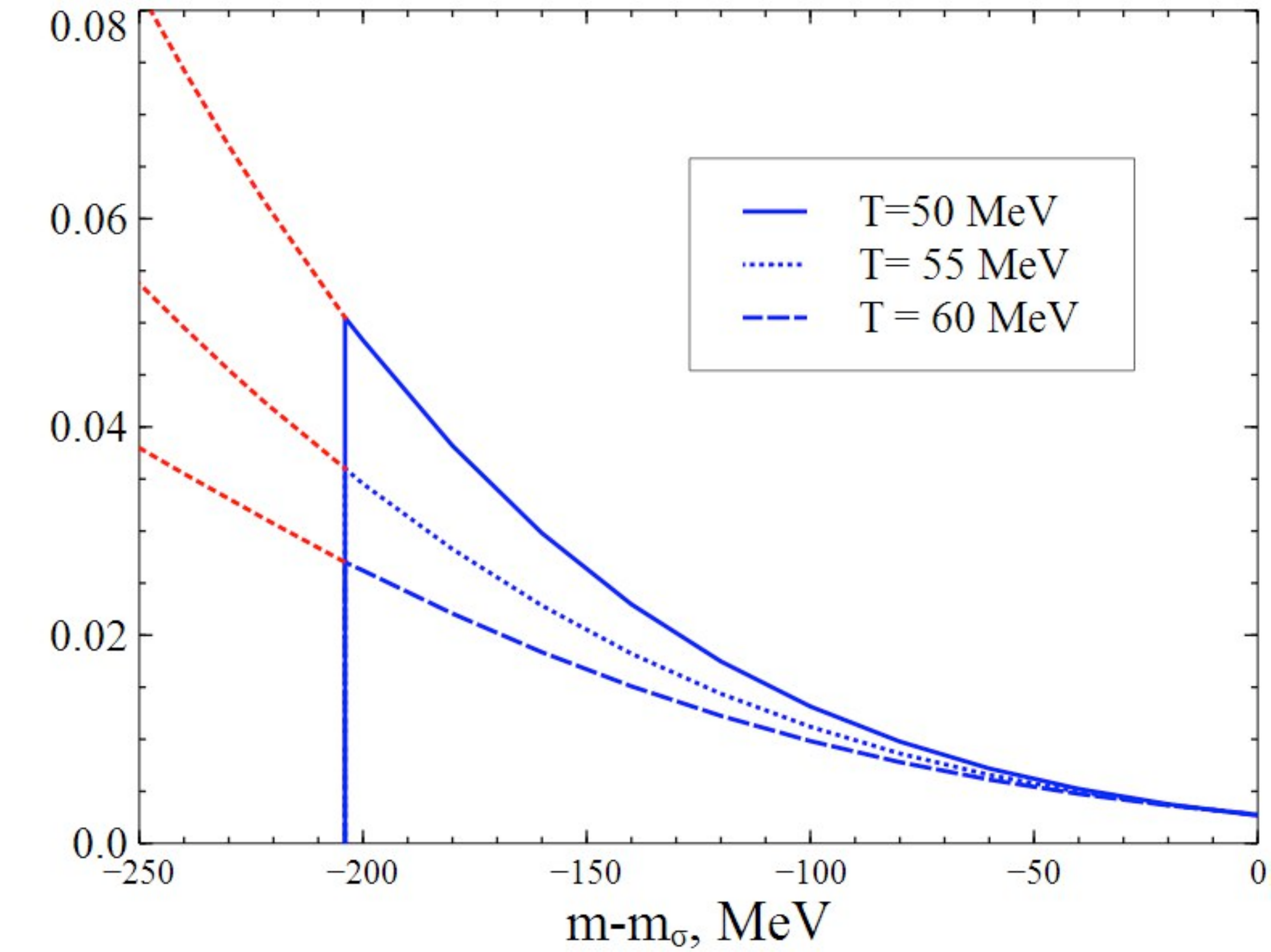}
}
 \caption{Temperature dependence of  the mass distribution  $f_k^G(m)/ \phi(m_\sigma, T)  $ (in units of $1/$MeV, see Eq. (\ref{EqIXn})) for a $\sigma$-meson  with the mass $m_\sigma = 484$ MeV, the width $\Gamma_\sigma = 510$ MeV  \cite{KABSigma:07} and $M_\sigma^{Th} = 2 \, m_\pi \simeq 280$ MeV.  In the upper panel the short dashed curves below the two pion threshold (vertical line at $m-m_\sigma = -204$ MeV)  show the mass attenuation  which does not contribute into the particle density (\ref{EqIXn}). From the lower panel one can see the effect of  wide resonance sharpening near the threshold, i.e. an appearance of a narrow peak in the resulting mass distribution on the right hand side of the threshold which resembles an  icy slide. For different temperatures this  mass attenuation   is shown by the solid, the short dashed
 and the long dashed curves.  The $\sigma$-meson  effective width   was found numerically from these mass attenuations:
 $\Gamma^{app}_\sigma (T=50\, {\rm MeV}) \simeq 62.5$ MeV,
 $\Gamma^{app}_\sigma (T=55\, {\rm MeV}) \simeq 71.5$ MeV and
  $\Gamma^{app}_\sigma (T=60\, {\rm MeV}) \simeq 82.5$ MeV.
 }
 \label{KABFig6}
\end{figure}

The momentum integral in (\ref{EqVIIIn}) can be written using the non-relativistic approximation $\phi(m, T) \simeq
\left[ \frac{m\, T}{2\, \pi} \right]^\frac{3}{2}\exp \left[   -\frac{  m } {T} \right] $. Then  to simplify the mass integration
in  (\ref{EqVIIIn})  one can make the full square in it from the  powers of  $(m_k -m)$ and get
\begin{eqnarray}
\label{EqIXn}
&&\hspace*{-0mm}
F_k (\sigma_k)  \equiv  g_k  \int\limits_{0}^\infty  d m \, f_k^G ( m)  \simeq   \tilde g_k    \int\limits_{0}^\infty  d m \,  \frac{\Theta\left(  m - M_k^{Th} \right)}{N_k (M_k^{Th})} \,  \exp \left[ - \frac{(\tilde m_k -m)^2}{2\, \sigma_k^2}  \right]
 \left[ \frac{m\, T}{2 \pi} \right]^\frac{3}{2}\exp \left[   -\frac{  m_k } {T} \right]  \,, ~
\end{eqnarray}
where the following notations for an effective resonance  degeneracy $\tilde g_k$   and  for an effective resonance mass $\tilde m_k$
\begin{eqnarray}\label{EqXn}
\tilde g_k     & \equiv &  g_k  \exp \left[  \frac{\sigma_k^2}{2 \, T^2} \right] = g_k  \exp \left[  \frac{\Gamma_k^2}{2 \,Q^2\, T^2} \right] \\
 \tilde m_k  & \equiv & m_k - \frac{ \sigma_k^2 }{T} = m_k - \frac{ \Gamma_k^2 }{Q^2\, T}
\label{EqXIn}
\end{eqnarray}
are used.  From Eq.  (\ref{EqIXn})
one can  see that  the presence of the width, firstly,  may  strongly  modify the degeneracy  factor $g_k$ and,
 secondly,  it may essentially shift the maximum of the mass attenuation towards  the threshold or even below it.
 There are two corresponding effects which we named as {\it the near threshold thermal resonance enhancement} and  {\it the near threshold resonance sharpening}.
 These effects formally appear due to the same reason as the  famous Gamow window for the thermonuclear reactions in stars \cite{KABGamow1, KABGamow2}:
 just above the resonance decay threshold
 the integrand in (\ref{EqIXn}) is a product of two functions of a virtual resonance mass $m$, namely, the Gaussian attenuation is an increasing
 function of $m$, while the Boltzmann exponent strongly decreases above the threshold. The resulting attenuation of their product has a maximum,
 whose shape, in contrast to the usual Gamow window,  may be extremely asymmetric due to the presence of the threshold.  Indeed,
 as one can see from Fig. \ref{KABFig6} the resulting mass attenuation of a  resonance may acquire  the form of  the sharp and narrow peak that is
 closely resembling
 an icy slide.
 Below we discuss   these two effects in some details.
Qualitatively the same effects appear, if in (\ref{EqIXn})  one uses the Breit-Wigner
 resonance mass attenuation instead of the Gaussian one (see below).

From the  definitions of  the effective resonance mass (\ref{EqXIn}) and the effective resonance degeneracy (\ref{EqXn})  one can see that the effects of their change are strong for $T \ll \sigma_k$. This can be clearly seen from  Fig. \ref{KABFig6}, which demonstrates both of the above effects at low temperatures  for  the $\sigma$-meson. A simple analysis shows that the effect of resonance sharpening is strongest, if the threshold mass is shifted to the convex part of the Gaussian distribution in (\ref{EqIXn}), i.e. for $M_k^{Th} \ge \tilde m_k$ or for the temperatures  $T$ below $T^+_k \equiv \frac{\sigma_k^2}{m_k - M_k^{Th}} \equiv \frac{\sigma_k}{\beta_k}$.
To  demonstrate the effect of the width sharpening we  list
a few typical examples for baryons  in the Table 1.   For $T < T^+_k$ and for $m > M_k^{Th}$ the Gaussian  mass distribution in (\ref{EqIXn}) can be safely approximated as $\exp \left[ - \frac{(\tilde m_k -m)^2}{2\, \sigma_k^2}  \right]  \approx \exp \left[ - \frac{(\tilde m_k - M_k^{Th})^2}{2\, \sigma_k^2} \, - \, \frac{(\tilde m_k - M_k^{Th})}{ \sigma_k^2}   (m - M_k^{Th}) \right]$. Now recalling the standard
definition of the width for the function $f (x) = \Theta(x)\, Const \, \exp\left[ - b \, x  \right]$, one obtains
the temperature dependent  resonance effective width near the threshold as
\begin{eqnarray}
\label{EqXIIn}
\Gamma_{kG}^{app} (T)  & \simeq  &  \frac{\ln(2)}{ \frac{1}{T} - \frac{\beta_k}{\sigma_k} } \equiv   \frac{\ln(2)}{ \frac{1}{T} -  \frac{1}{T^+_{kG}} }    \,,
\end{eqnarray}
 since for such a distribution function $f(x)$ one gets $f(\ln(2)/b) = f(0)/2$. Note that in evaluating (\ref{EqXIIn}) we neglected the additional $m^{1.5}$-dependence in (\ref{EqIXn}), but one can readily check that numerically such a correction is small.  The rightmost column in  Table 1
 demonstrates that  Eq. (\ref{EqXIIn}), indeed, provides an accurate estimates for $T < T_{kG}^+$.  The results of  Table 1   also justify the usage of $\sigma$-meson and the field theoretical models based on the well known $\sigma$-model for temperatures well  below $T^+_\sigma \simeq 92$ MeV.  Of course, the present approach which is developed for the chemical FO stage, when the inelastic reactions except for resonance decays are ceased to exist, cannot be applied for earlier stages of heavy ion  collisions. However, here we would like to stress  that an inclusion of the large width of $\sigma$-meson in the field theoretical models of the strongly interacting matter equation of state is very necessary. From the above analysis one can see that the large width inclusion can generate some new important physical effects like the wide resonance sharpening in a thermal medium.

\begin{table*}[!]
%
\noindent
\caption{
The parameters of several hadronic resonances together with their  decay  channels  that are used to determine the quantities $\beta_k$ and $T_{kG}^+$. The last two columns show the corresponding effective width at temperature $T=50$ MeV found, respectively, numerically from Eq. (\ref{EqVIIIn}) and analytically  from Eq. (\ref{EqXIIn}), when it can be applied.
}
 \vskip3mm\tabcolsep10.7pt
\begin{tabular}{|c|c|c|c|c|c|c|c|c|c|c|c|c|}
\hline
     Hadron &  $m_k$  & $\Gamma_k$ &   Decay    &   $M_k^{Th}$  & $\beta_k$ & $T_{kG}^+$   &  exact
     $\Gamma_k^{app}$  & approx. $\Gamma_k^{app}$ \\
                   &  (MeV)       &            (MeV)          &   ~ channel ~    &        (MeV)      &  &        (MeV)  &       (MeV) &             (MeV) \\\hline
         $\sigma$-meson    &         484        &     510    &      $\sigma \rightarrow \pi \pi$   &    280  & 0.942 & 91.9 &
         62.5 &  67.3 \\  \hline
          $P_{33}$     &   1232  &    120   &    $\Delta \rightarrow \pi N$    &    1080  & 2.98 & 11.6 &  43.5 & N/A  \\  \hline
          $P_{11}$     &   1440  &    350   &    $N \rightarrow \pi N$    &    1080  & 2.42 & 38.74 &  129.5 & N/A  \\  \hline
          $P_{33}$     &   1600  &    350   &    $\Delta \rightarrow \pi \Delta$    &    1372  & 1.53 & 50.4  & 68.7 & 80.8 \\  \hline
          $P_{33}$     &   1600  &    350   &    $\Delta \rightarrow \pi N$    &    1080  & 3.5 & 30.3 &  280. & N/A \\  \hline
            $G_{17}$     &   2190  &    500   &    $\Delta \rightarrow \rho N$    &    1710  & 2.26 & 57.8  & 74.6 & 81.8 \\  \hline
\end{tabular}
\label{table1}
\end{table*}

\begin{figure}[t]
\centerline{\includegraphics[width=8 cm]{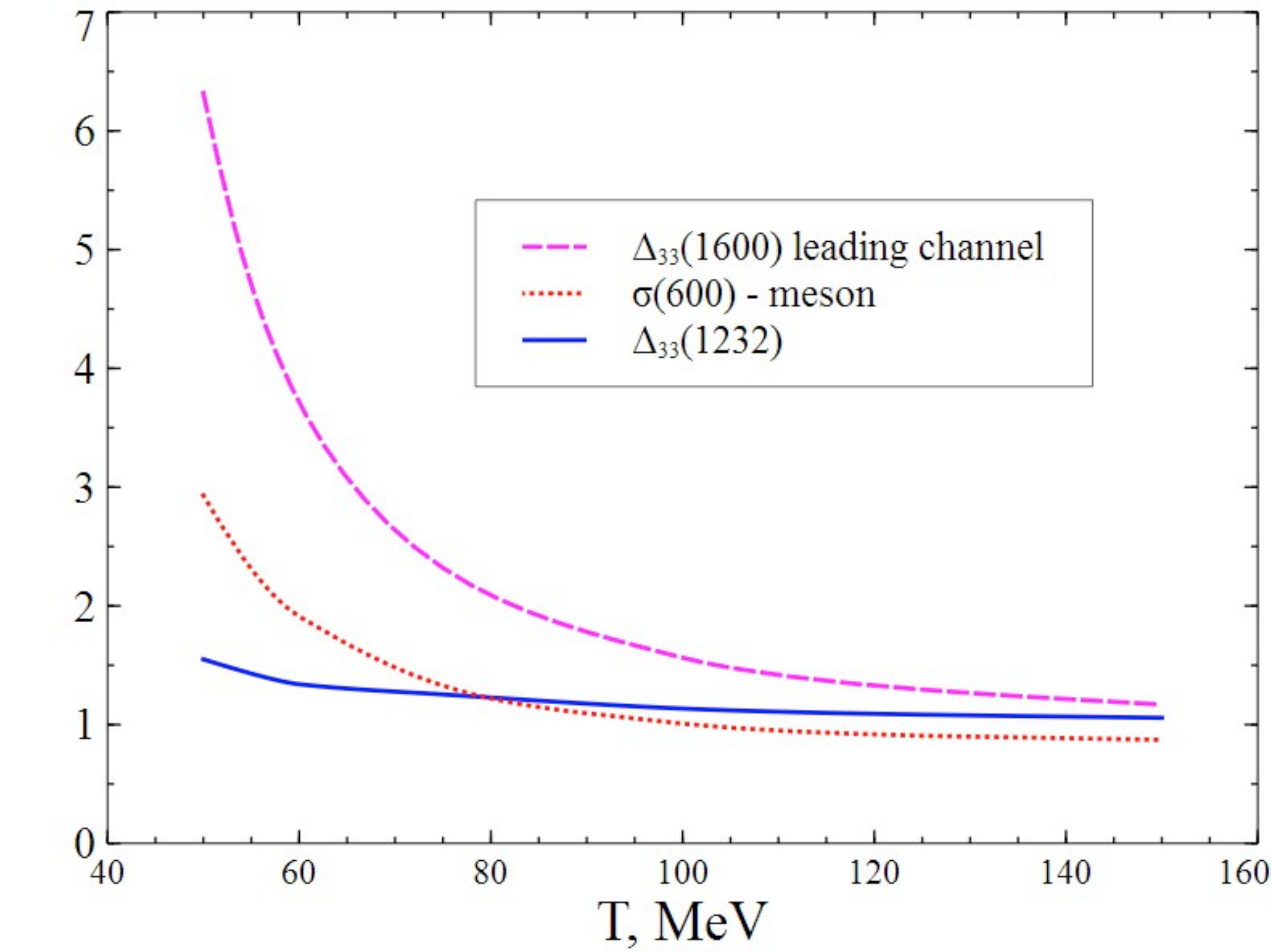}}   \vspace*{2.2mm}
\centerline{\includegraphics[width=8 cm]{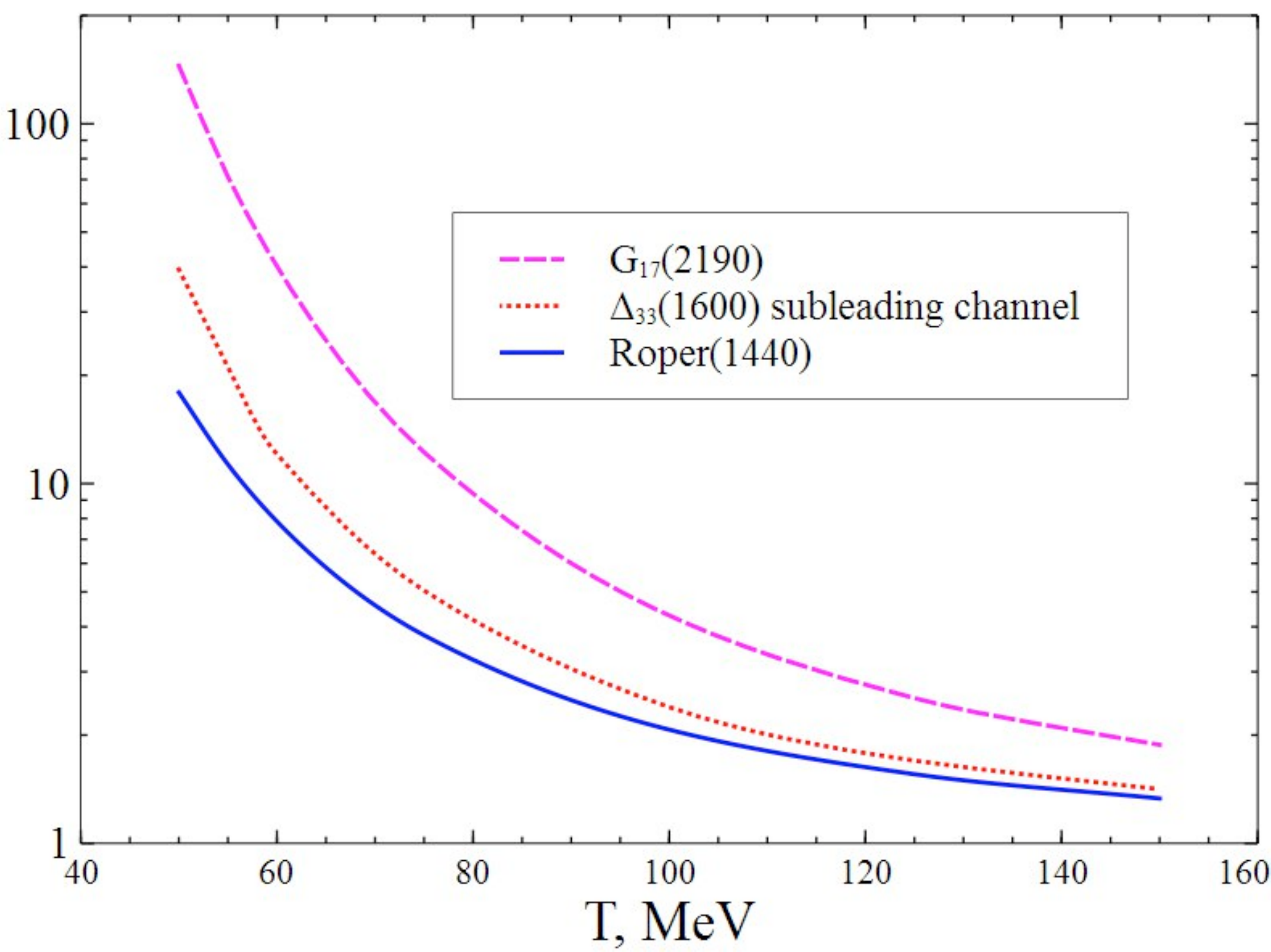}
}
 \caption{Temperature dependence of  the resonance  enhancement factor  $R_G = \frac{F_k(\sigma_k)}{g_k\, \phi (m_k,T)}$ for the HRGMG. The resonance  enhancement factor  $R_G$  is shown for the hadronic resonance decays given in Table I.  At low temperatures the enhancement factor  of  wide resonances can be huge.
  }
 \label{KABFig7}
\end{figure}

From  Fig. \ref{KABFig7} one can see that the resonance enhancement  can be, indeed,  huge for wide ($\Gamma \ge  450$ MeV) and medium wide ($\Gamma \simeq 300-400$ MeV) resonances.
This effect naturally explains an enhancement of  hadronic pressure  and energy density
at chemical FO compared to the zero width case (see Fig. \ref{KABFig3}). It is appropriate here to define the apparent width for the Breit-Wigner mass attenuation.    Replacing in  (\ref{EqVIIIn})  the Gaussian mass attenuation by the Breit-Wigner one,    instead the  function $f_k^G ( m)$ in (\ref{EqIXn}) we get
\begin{eqnarray}
\label{EqXIIIn}
&&\hspace*{-0mm}
f_k^{BW} ( m)   \simeq   \frac{\Theta\left(  m - M_k^{Th} \right)}{N_k^{BW} (M_k^{Th})}\cdot
 \frac{m^{- B}}{(m_k -m)^2 + \frac{\Gamma_k^2}{4}} \, \left[ \frac{m\, T}{2 \pi} \right]^\frac{3}{2}\exp \left[   -\frac{  m } {T} \right]  \,, 
 \end{eqnarray}
 \begin{eqnarray}
 \label{EqXIVn}
&&\hspace*{-4mm}N_k^{BW} (M_k^{Th})   \equiv   \int\limits_{0}^\infty  d m \,  
 \frac{\Theta\left(  m - M_k^{Th}  \right) \, m^{- B}}{(m_k -m)^2 + \frac{\Gamma_k^2}{4}}  \,.
\end{eqnarray}
Here we introduced an additional mass dependence factor $m^{- B}$ for convenience ($B = const$). Calculating the  mass derivative  of $\ln(f_k^{BW} ( m))$  for $m > M_k^{Th}$ one finds:
\begin{eqnarray}
\label{EqXVn}
&&\hspace*{-0mm}
\frac{d }{d m} \ln (f_k^{BW} ( m))  \simeq    -\frac{1}{T}~ -  ~\frac{2(m -m_k)}{(m_k -m)^2 + \frac{\Gamma_k^2}{4}}  ~
 +~ \frac{\frac{3}{2}-B}{m} \,.
\end{eqnarray}
Similarly to the case of the Gaussian width,  from (\ref{EqXVn}) one can find an apparent width for the Breit-Wigner case  as
\begin{eqnarray}
\label{EqXVIn}
&&\hspace*{-0mm}
\Gamma_{kBW}^{app}   \simeq   - \frac{\ln(2)}{
\frac{d }{d m} \ln (f_k^{BW} (M_k^{Th}+0 )) } =    \frac{\ln(2)}{\frac{1}{T}~ - \frac{1}{ T^+_{BW}} } \,,\\
\label{EqXVIIn}
&&\hspace*{-0mm}\frac{1}{ T^+_{kBW}}  \simeq    \frac{2\, \beta_k}{\sigma_k \left(\beta_k^2  + \frac{Q^2}{4} \right)}
 + \frac{\frac{3}{2} - B}{M_k^{Th}} \equiv  \frac{2}{T^+_{kG} \left(\beta_k^2  + \frac{Q^2}{4} \right)}
 + \frac{\frac{3}{2} - B}{M_k^{Th}}\,,
\end{eqnarray}
where in the last step of evaluation we used the same notations as in Eq. (\ref{EqXIIn}). From Eq.
(\ref{EqXVIIn}) one can deduce two conclusions.
First, for $M_k^{Th} \gg T$ the last term on the right hand side of
(\ref{EqXVIIn}) can be safely neglected, if $O(\frac{3}{2} - B) \sim 1$.
This fact shows that power-like  deformations of the Breit-Wigner attenuation cannot affect the apparent
width (\ref{EqXVIn}).  This is the main  reason why the tiny deformations of the resonance
mass attenuation discussed earlier with respect to claims of Ref.  \cite{KABFriman} should not be taken
seriously, since they are much weaker than the effect of a thermal environment clearly seeing  in Eqs.
(\ref{EqXIIn}) and (\ref{EqXVIn}).

Second, for $\beta_k \ge 1$ one finds that ${2}{ \left(\beta_k^2  + \frac{Q^2}{4} \right)^{-1}} \simeq \frac{2}{ \left(\beta_k^2  + 1.4 \right)} < 1$. From the last inequality of finds two important inequalities
\begin{eqnarray}
\label{EqXVIIIn}
\frac{1}{ T^+_{kBW}} & <  &  \frac{1}{ T^+_{kG}} \quad \Rightarrow  \quad \Gamma_{kBW}^{app} (T) ~< ~\Gamma_{kG}^{app} (T) \,.
\end{eqnarray}
The inequalities  above are valid for $T <  T^+_{kG}$. They show that   
compared to the Gaussian width  parameterization 
 the range of mass states which contribute into the pressure defined by the Breit-Wigner mass attenuation of wide resonances is more close to the threshold.  In order to demonstrate how
an apparent width of wide resonances behaves at higher temperatures, it is instructive to study in some details the enhancement factor for Breit-Wigner and Gaussian attenuations.
For this purpose it is convenient to rewrite the enhancement factor  $R = \frac{F_k(\sigma_k)}{\phi (m_k,T)}$ of heavy resonances  in the non-relativistic approximation $(m_k \gg T)$ as 
\begin{eqnarray}\label{EqXXVI}
&&R_{G} (\beta, t)   = \int\limits_{-\beta}^\infty   \frac{ d x \left[1 + s\, x  \right]^\frac{3}{2} \,  \exp \left[ -\frac{x^2}{2} - \frac{x}{t}  \right] }{ \int\limits_{-\beta}^\infty  {dy}\,{ \exp \left[ -\frac{y^2}{2} \right]  } } \, , \\
&&R_{BW} (\beta, t)  = \int\limits_{-\beta}^\infty  \frac{d x \left[1 + s\, x  \right]^\frac{3}{2} \, \exp\left( - \frac{x}{t}\right) }{\left( x^2 + \frac{Q^2}{4} \right)  \int\limits_{-\beta}^\infty \frac{d y}{ \left( y^2 + \frac{Q^2}{4} \right)  } } \, ,
\label{EqXXVII}
\end{eqnarray}
where for the resonance of $k$-th sort the  parameters are defined as follows:  $s$  stays for  $s_k  \equiv  \frac{\sigma_k}{m_k}$, while  $\beta$ stays for $\beta_k$ and the  reduced  temperature $t$ is given in terms of the Gaussian resonance width $\sigma_k$ as $t_k \equiv T/\sigma_k$. The $x$ and $y$ integrations in  Eqs. (\ref{EqXXVI}) and  (\ref{EqXXVII}) are performed over the dimensionless  variable $(m -m_k)/\sigma_k$.

\begin{figure}[t]
\centerline{\includegraphics[width=8 cm,height=7cm]{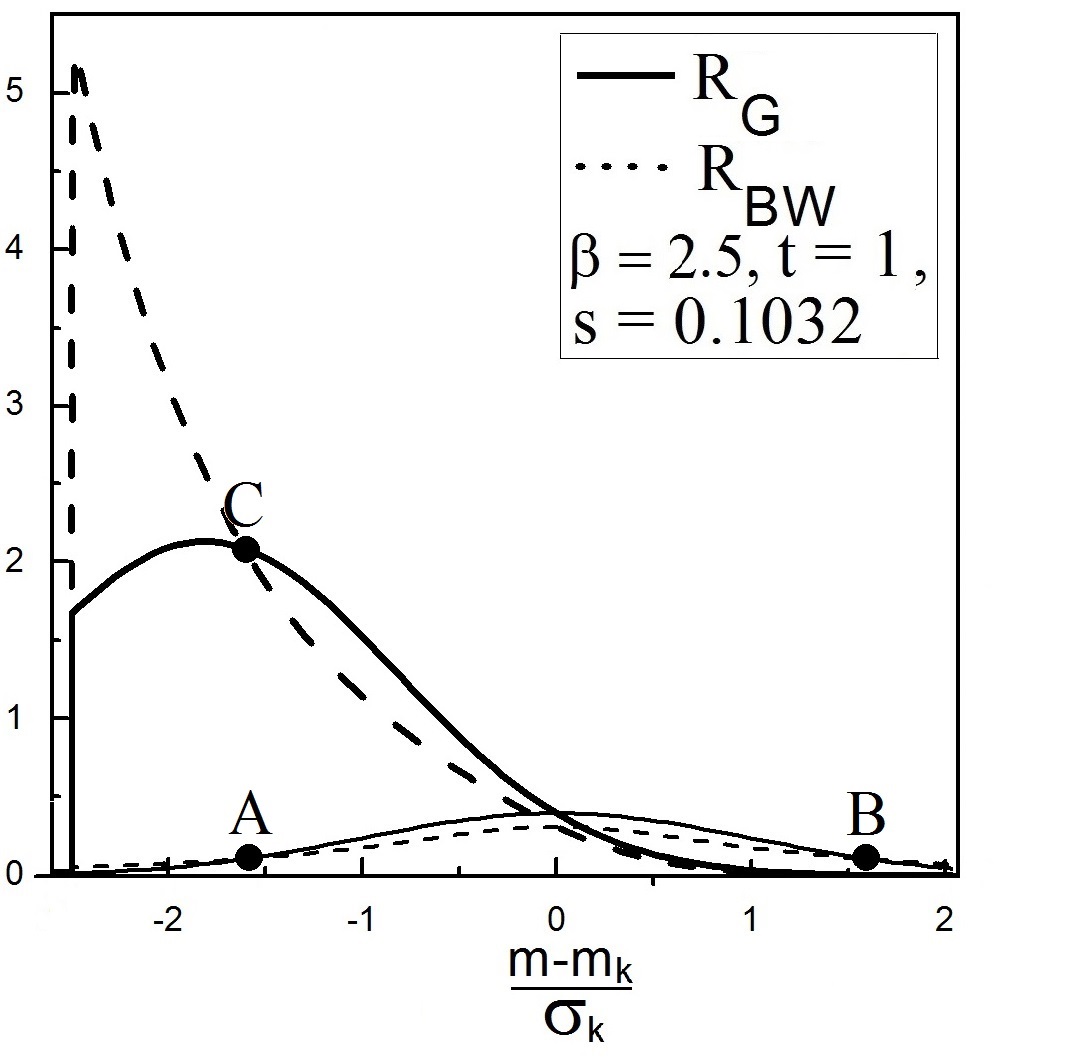}}   \vspace*{2.2mm}
\centerline{\includegraphics[width=8 cm,height=7cm]{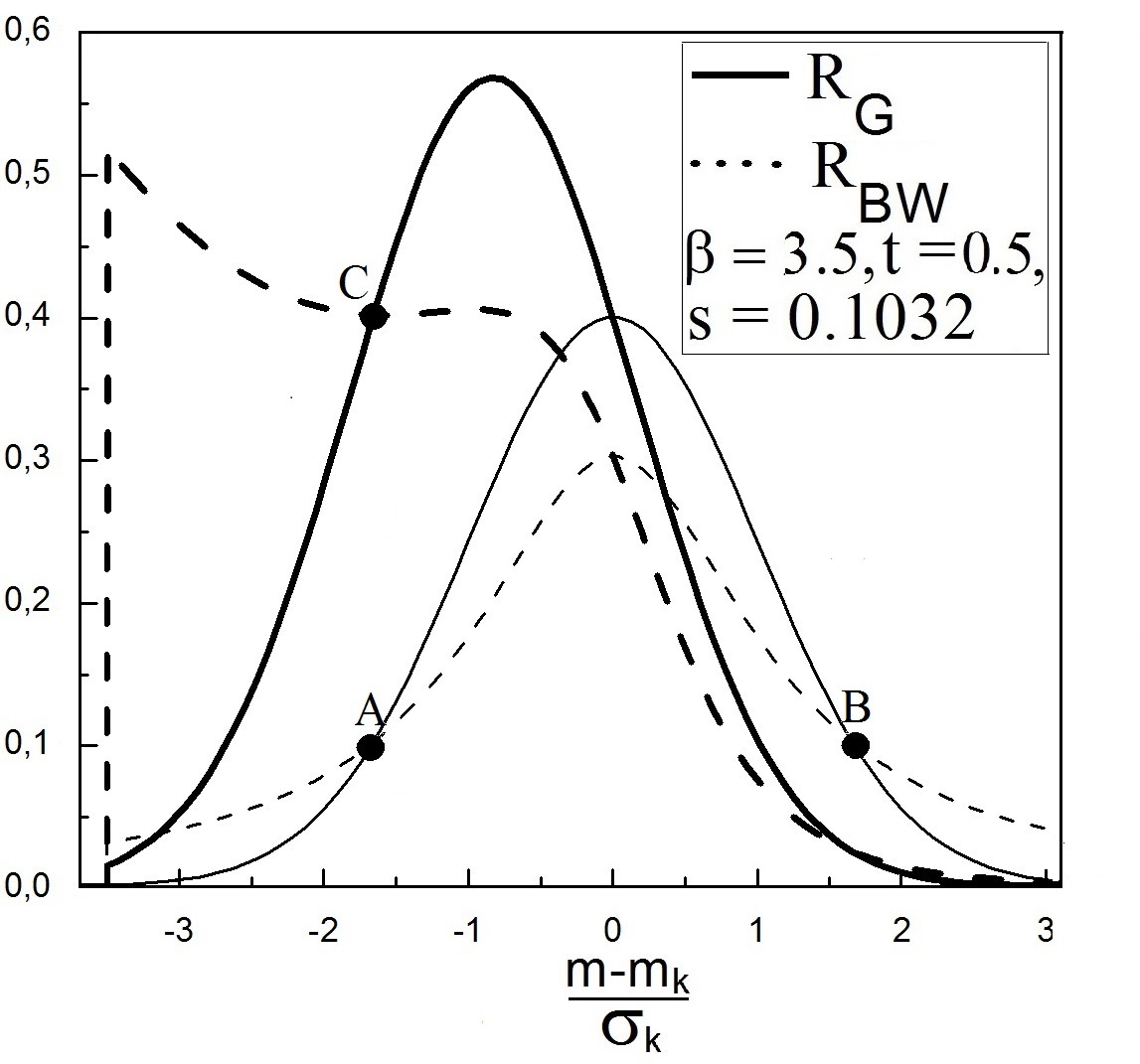}
}
 \caption{Mass  attenuations of  the resonance  enhancement  factors $R_{G}$  and $R_{BW}$
 are  compared  for two values of the reduced temperature $t =T/\sigma_k$. Upper (lower)  panel corresponds to $t =0.5$ ($t =1$).   The 
 thin solid curve represents the Gaussian mass attenuation, while the thin dashed curve shows the Breit-Wigner one.
 The thick  curves correspond to the integrands staying in  Eqs. (\ref{EqXXVI}) (solid) and (\ref{EqXXVII}) (dashed). 
 The intersection points A, B and C are discussed in the text.}
 \label{KABFig8}
\end{figure}

\begin{figure}[t]
\centerline{\includegraphics[width=8 cm,height=7cm]{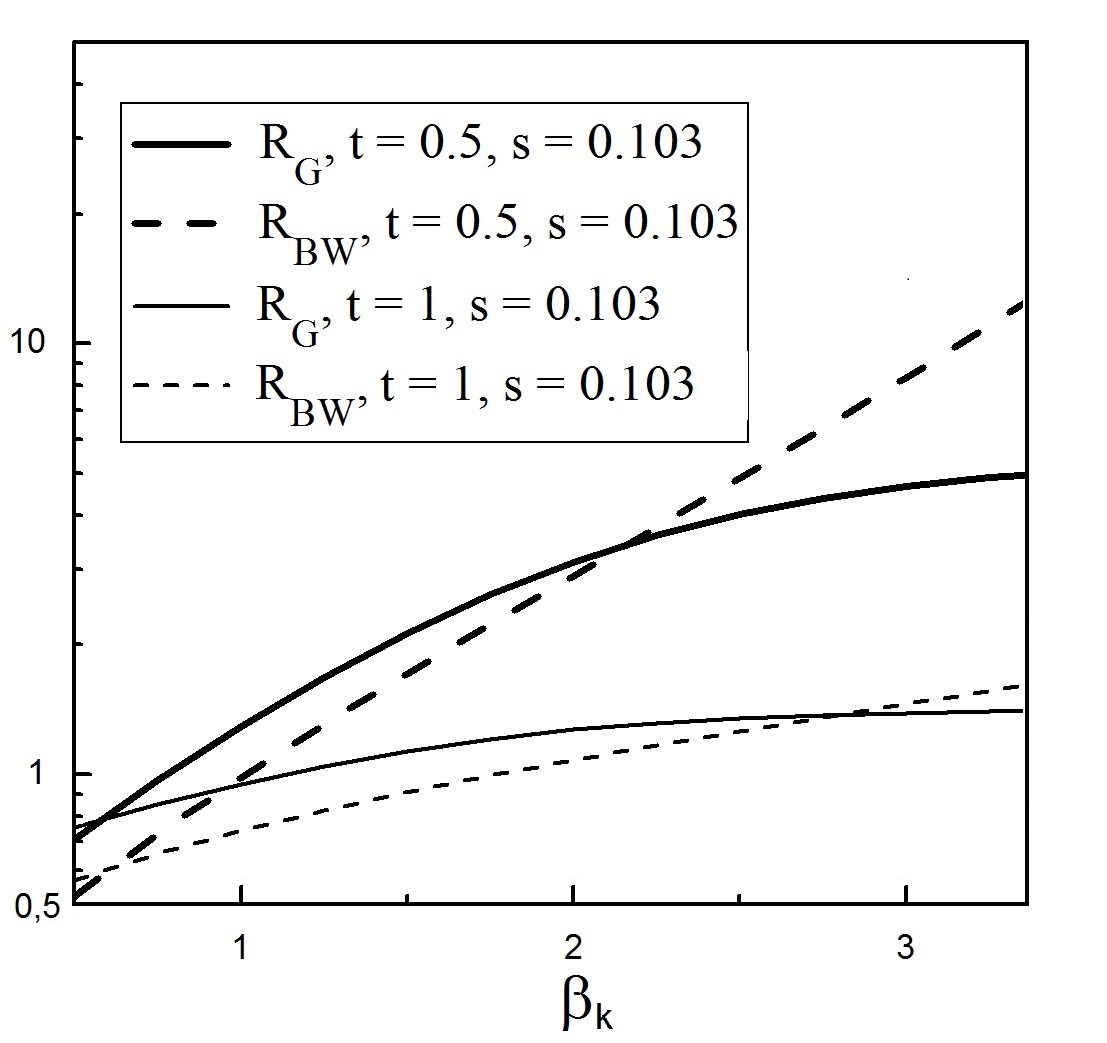}}   
 \caption{$\beta_k$  dependence of  the resonance  enhancement factor $R$  is shown for two values of the reduced temperature $t=T/\sigma_k$. Thick (thin)  curves  correspond  to $t =0.5$ ($t =1$).   The 
solid curves represent the Gaussian mass attenuation of resonances, while the dashed curves show  the results for the  Breit-Wigner one.
}
 \label{KABFig9}
\end{figure}

The enhancement factors (\ref{EqXXVI}) and  (\ref{EqXXVII}) demonstrate a strong dependence on 
$\beta$ and $t$, and a weak one on the variable $s$.  Therefore,  in Figs. \ref{KABFig8} and \ref{KABFig9} we depict the results for some typical values of  $\beta_k$ and $t_k$, while fix 
the $s$ value to $s = s_k \simeq 0.1032$ which corresponds to the Roper resonance.
Actually, the parameters chosen for the upper and the lower  panels of  Fig. \ref{KABFig8}   are, respectively,  very close  to the parameters  of  the Roper  resonance (compare $\beta =2.5$ in  Fig. \ref{KABFig8}  and $\beta_{Roper} \simeq 2.42$) and the $P_{33}$ resonance (compare $s=0.1032$ in Fig. \ref{KABFig8} and $s_{P_{33}} \simeq 0.093$)   which decays into pion and nucleon (see Table 1).   Therefore, 
in case of the  Roper  resonance an actual temperature of the upper panel of  Fig. \ref{KABFig8} is about 149 MeV, while for the $P_{33}$ resonance with $\beta_{P_{33}} \simeq 3.5$ its lower panel 
corresponds to the temperature of about 75 MeV.  

As one can see from the upper panel of Fig. \ref{KABFig8}  the  Breit-Wigner mass attenuation is more narrow than the Gaussian one, while the lower panel of this figure corresponds to an opposite case.  Nevertheless, as it is clear from Fig.  \ref{KABFig9} both of the cases correspond to the inequality $R_{BW} > R_G$.  In order to understand this  inequality 
we note that the normalized Breit-Wigner and Gauss mass distributions used in Eqs. (\ref{EqXXVI}) and  (\ref{EqXXVII}) get equal  for $x_k^\pm  \simeq \pm (1.6 \pm 0.1)$ depending on the value of $\beta_k$, where $x$ denotes  the dimensionless variable $\frac{m-m_k}{\sigma_k}$. These intersection points of two normalized distributions are denoted by the points A and B in Fig. \ref{KABFig8}.  Now it is clear that, if the threshold is located between $x=0$ and $x= x_k^-$, i.e.  for $-\beta_k \ge   x_k^-$, then the enhancement factor of the Gaussian mass attenuation $R_G$ is larger than $R_{BW}$. A similar situation exists, if the threshold is not too below $x= x_k^-$, i.e.  for $-\beta_k  \leq x_k^-$. However, if  the threshold is located somewhat   away from the intersection point $x= x_k^-$, i.e. for $-\beta_k \ll x_k^-$, then the Breit-Wigner mass attenuation is essentially enhanced near the threshold by the Boltzmann exponential, as one can see from the both panels of Fig. \ref{KABFig8}. Evidently, that effect gets stronger for lower values of  temperature and the dashed lines in  Fig. 
\ref{KABFig9} clearly demonstrate an exponential dependence for the  Breit-Wigner enhancement factor $R_{BW} \sim \exp [ \beta_k/t_k]$ for $\beta_k > 2 $.

Also it is important that  the Breit-Wifner  enhancement factor of the narrow and very narrow  resonances is  larger than the Gaussian one, if the threshold is located somewhat away from the mean resonance mass $m_k$. For example, at   the  temperature 100 MeV   for the $\omega (783)$-meson which with  a small width  decays into three pions these factors are $R_{BW} \simeq 1.034$ and  $R_G = 1$.  A more dramatic difference at this temperature  one finds for the $\rho(770)$-meson which decays into two pions:
$R_{BW} \simeq 1.79$ and  $R_G  \simeq  1.14$. Thus, for  $-\beta_k \ll x_k^-$, which is the case for  
the $\omega (783)$-meson ($\beta_\omega \simeq 89.6$) and for the $\rho(770)$-meson 
($\beta_\rho \simeq 7.54$), the   Breit-Wigner enhancement factor $R_{BW}$ exceeds  the Gaussian one $R_{G}$, i.e. $R_{BW} > R_{G}$. 

It is also necessary to note that, if  for a given resonance an inequality  $R_{BW} < R_{G}$ takes place, then  usually the Gaussian  enhancement factor  does not exceed 40 \% of $R_{BW}$  and very seldom it  exceeds  50 \%, while  for $T < 100$ MeV  we found  many examples that an opposite inequality for these enhancement factors can be,  in fact,  replaced by the following one $R_{BW} \gg  R_{G}$. 
The  typical examples of such a behavior are  given by the tick curves  in Fig.  \ref{KABFig9}. 
The discussed properties of the enhancement factors allow one to naturally explain the fact, that the chemical FO pressure of  the HRGMBW can be twice larger  than the one of  the HRGMG (see an upper panel of  Fig. \ref{KABFig3}), although the relative difference of the corresponding  chemical FO temperatures and baryonic chemical potentials is less, than 6 \%.
Thus, the presence of many resonances (even very narrow ones!), whose decay thresholds are far away from the peak of the resonance  mass attenuation, i.e.   for  $-\beta_k \ll x_k^-$, leads to a strong enhancement  of  the HRGMBW pressure  compared  to the HRGMG pressure, while the latter  is also enhanced  compared to  the  HRGM0 pressure.
 For the energy density such a conclusion is not obvious,  since in addition  one has to take into account
for the density of states with a given mass.

The first important result  from this  analysis is that there is no sense to discuss the mass spectrum of
hadronic resonances, empirical or Hagedorn, without a treatment of  their width.
Clearly, the same is true
for  the QG bags which, according to the FWM \cite{FWM,FWMb},  are heavy and wide  resonances
with mass $M_B$ larger than $M_0 \simeq 2.5$ GeV and with  the mean width of the form $\Gamma_B \simeq \Gamma_0 (T) \left[  \frac{M_B}{M_0}  \right]^\frac{1}{2}$,  where  $\Gamma_0 (T)$ is a monotonically increasing function  of $T$ and  $\Gamma_0 (T=0) \in [400; 600]$ MeV.
This range of   $\Gamma_0 (T=0)$  values corresponds to the pseudo-critical  temperature $T_{pc} \simeq 170-200$ MeV \cite{FWM,FWMb} for vanishing baryonic density. The value $\Gamma_0 (T=0)=400$ MeV   is well consistent with the results of the
present days lattice QCD thermodynamics \cite{KABKarsch,KABWupBuD:2009}, but there is no
guaranty that the lattice QCD data will not change in the future. Therefore, below  we consider the whole range of
values for the width $\Gamma_0 (T)$ analyzed in \cite{FWM,FWMb}.

There are two interesting features of  QG bags which are related to the above treatment. Thus, from the results of  \cite{FWM,FWMb} and from Eq. (\ref{EqXIIn})  one can find   the  temperature $T^+_B$  for the meson-like  QG bags  as
\begin{eqnarray}
\label{EqXXX}
&&\hspace*{-0mm}T^+_B \simeq  \frac{ \Gamma_0^2 (T) }{Q^2\, M_0 \left(1 - \xi_B  \right) } \simeq
\frac{1}{\left(1 - \xi_B  \right) }\,  \times \left\{
\begin{array}{ll}
11.5-26\,\, \rm MeV  \,,
&
{\rm if} ~ ~\Gamma_0  (0) \simeq 0.4-0.6\,\, {\rm GeV}\,,  \\
46-104\,\, \rm MeV  \,,
&
{\rm if} ~ ~\Gamma_0 (90) \simeq 0.8-1.2\,\, {\rm GeV}  \,,  ~ \\
140-315\,\, \rm MeV  \,,
&
{\rm if} ~ ~\Gamma_0  (170) \simeq 1.4-2.1\,\, {\rm GeV}  \,,
\end{array}
\right. \label{EqXIXn}
\end{eqnarray}
where  $\xi_B \equiv \frac{M_B^{Th}}{M_B}$ denotes the ratio of the leading threshold mass  $M_B^{Th}$ of the bag to  its  mean mass  $M_B$.
In (\ref{EqXIXn}) the values of temperature $T$ which are given in an expression $\Gamma_0  (T)$ are  measured in MeVs.  Our  justification to apply  (\ref{EqXIXn}) to the meson-like QG bags at non-zero baryonic densities was given above.

 Clearly,  for different bags the range of  $\xi_B$ value can be  between 0 and 1.
Therefore,  according to above results  the bags with $\xi_B \rightarrow  1$ should have been essentially enhanced and
sharpened as the ordinary resonances. Moreover,  according to (\ref{EqXIIn}) in this case for $T  \ll T_B^+$ the meson-like QG bags should have had a small width $\Gamma_B^{app} \simeq \frac{T \, T_B^+}{T_B^+ - T} \ln(2)$ and, hence, such QG bags should have long life-time or, in other words,
there is a chance to observe such  QG bags! We remind that the reason why such bags are not observed in the experiments is naturally explained by the FWM  \cite{FWM,FWMb}:
it is due to  {\it the subthreshold suppression} (for more details see a discussion after Eq. (42) in \cite{FWMb}).

On the other hand Eq. (\ref{EqXXX})  shows that the only hope to observe the QG bags exists, if $\xi_B \rightarrow 1$. Then for chemical FO temperatures much below $T^+_B$ such bags could have sufficiently long  eigen lifetime of about $\tau_B \sim \frac{1}{\Gamma_B^{app}} \simeq \frac{T_B^+ - T }{T \, T_B^+\, \ln(2)} \le  \frac{1}{T \,  \ln(2)} $.
Substituting  $T \simeq 0.5 \, T_H \in [80; 90]$ MeV in the last inequality and using the estimate of  Eq. (\ref{EqXIXn}) for $T=90$ MeV with $\xi_B = 0.9$, one finds the most optimistic estimates for the QG bag eigen lifetime as  $\tau_B \le 3.3 \pm 0.3 $ fm/c.
On the other hand  for  $\xi_B = 0.9$ and  $T=140$  MeV from  (\ref{EqXIIn}) and (\ref{EqXIXn})
one finds that  $T_B^+ \simeq 1200$ MeV and $\Gamma_B^{app}  \simeq  100-120$    MeV.
These estimates allow us to make the second important conclusion  that {\it   an  appearance  of   sharp  resonances (mesonic  or/and  baryonic) with the apparent width being 
in the interval between 50 to 120 MeV}  at the chemical FO  temperatures   $T_{QGB} \simeq  85 -140$ MeV  {\it that have  the mass
above 2.5 GeV and that are absent in the tables of particle properties would be  a clear signal of the  QG bag formation.}
Their possible appearance at chemical FO as metastable states of  finite systems created in relativistic nuclear collisions  is justified by the FWM
\cite{FWM,FWMb}.
At higher temperatures such QG bags can be formed too, but their apparent width is larger and their  enhancement  is less pronounced. The limiting values of  $\xi_B$ at $T$ for which the effect of resonance sharpening can exist is determined by the relation $\Gamma_0^2 (T)/T \ge Q^2\, M_0 \left(1 - \xi_B  \right)$. From the latter inequality  one can see that the condition $\xi_B \rightarrow 0.9$ can be relaxed, but  in this case the temperature of chemical FO  gets higher.

In addition one  has to account for  the statistical probability of  the QG bags appearance at a given temperature $T$. Relatively to the nucleon  the thermal   probability of the QG bag of mass $M_B$ is about $W = \left[ \frac{M_B}{M_N} \right]^{1.5} \exp \left[ \frac{(M_N-M_B)}{T} \right] \, R_B(T)$, where $M_N \simeq 940$ MeV is the nucleon mass and  $R_B(T)$ is the resonance enhancement factor in a thermal medium.  For $T= 140$ MeV and $M_B = M_0 \simeq 2.5$ GeV one gets $W_B \simeq 3.85 \cdot 10^{-5} R_B$. Above it was shown  that for such temperatures  the typical  resonance apparent  width values  are about $\Gamma_B \simeq 100 -120$ MeV while the typical values of the resonance enhancement factor can be estimated as  $ R_B \simeq 10-100$ for $\beta_B \simeq 2.5-4.5$.     Therefore, compared to a nucleon  the relative thermal  probability of such QG bags is about $W_B \simeq 3.85 \cdot (10^{-4}- 10^{-3})$, which is  essentially larger than the relative probability of the $J/\psi$ meson  $W_{J/\psi} \simeq 1.19 \cdot 10^{-6}$ at the same temperature. Note that the chemical FO temperature $T \simeq 140$ MeV corresponds to the highest SPS energy of collision at which the $J/\psi$ mesons are safely measured.
Therefore, these estimates  give us a hope that the decays of  meson-like QG bags can, in principle,  be
measured in the energy range  $\sqrt{s_{NN}} \simeq 4.3-6$ GeV.

\section{Conclusions}

Here we  developed  the HRGM with the Gaussian mass attenuation of hadronic resonances.
A successful fit of the particle yield ratios allowed us to elucidate  an unprecedented    jump of the number of effective degrees of freedom existing  in the narrow energy range $\sqrt{s_{NN}} =$ 4.3--4.9 GeV.
It is remarkable that all realistic versions of  the HRGM analyzed here demonstrate the same behavior.
Therefore,  the developed concept  was  named the non-smooth chemical FO.
An effort to explain  the non-smooth chemical FO  led us to a conclusion that these irregularities
can be  related to the  QG bags formation at $\sqrt{s_{NN}} =$ 4.3--4.9 GeV.
Here we give some arguments based of the FWM framework that the found  irregularities, especially
a  jump in the number of effective degrees of  freedom,  can serve as
the  signals of the QG bags formation.

Also we analyzed the behavior of  wide resonances in a thermal environment, calculated their apparent width for the Gaussian and Breit-Wigner mass attenuations  and   found  two new effects occurring, if  the chemical FO  temperature is small  compared to the resonance width: {\it the near threshold thermal resonance enhancement} and  {\it the near threshold resonance sharpening}.
As we discussed, these effects formally appear due to the same reason as the  famous Gamow window for the thermonuclear reactions in stars.
The found effects allowed us to naturally explain the fact that the HRGM which accounts for the finite widths of hadronic resonances
generates higher   pressure than the one with a zero resonance width. 
As an important application of the found effects we studied   an apparent width of  the $\sigma$-meson for the Gaussian mass attenuation. A subsequent  analysis showed that for the temperatures well below $92$ MeV the $\sigma$-meson can be rather narrow and it  has  an apparent  width of about 50 to 70 MeV. Thus, accounting for the $\sigma$-meson large width in a thermal medium allows us  to justify the usage of the $\sigma$-like field theoretical models for the strongly interacting matter equation of state for temperatures well below $92$ MeV.

The new effects are thoroughly compared for the  Gaussian and the Breit-Wigner mass attenuations and 
it is shown that for the  same conditions the  Breit-Wigner width parameterization 
enhancement factor
can be essentially larger than the Gaussian one,  if the decay thresholds of a given resonance  are far 
away from the resonance peak. 
Such a result helps one to understand the reason of why 
the chemical FO pressure of  the HRGMBW can be twice larger  than the one of  the HRGMG, although the relative difference of the corresponding  chemical FO temperatures and baryonic chemical potentials does not exceeds  6 \%.
In the present work  a question of the correct parameterization of the resonance mass attenuation 
to be used for the phenomenological applications  is addressed to the microscopic models, which
at the moment do not provide us with the satisfactory prescriptions. 

Finally, applying   these  new effects  to the QG bags  we argued that the most optimistic chance  to find  the meson-like QG bags  experimentally  would be related to their sharpening and enhancement  by  a thermal medium.  If this is the case, then the QG bags may appear directly or in decays as narrow  resonances of the apparent  width about 50-120 MeV which  have  the  mass about or above 2.5 GeV and  which   are absent in the tables of elementary particles.

The practical conclusions that can be drawn  out of  these findings maybe important for planning experiments at the  FAIR (GSI) and at  the Nuclotron (JINR) facilities. 
As we argued here, it is possible that the energy range associated with the onset of the  quark-gluon-hadron mixed phase  does not correspond to the Strangeness Horn energy $\sqrt{s_{NN}} \simeq 7.6$ GeV, but   such an onset  should  be  searched  at the collision energies $\sqrt{s_{NN}} \simeq 4.3-4.9$ GeV.

\vspace*{3mm}

\noindent
{\bf Acknowledgments.} The authors are thankful to Z. Fodor, A. B. Larionov and  I. N. Mishustin  for valuable comments.
K.A.B., A.I.I.  and G.M.Z.  acknowledge  a  support
of  the Fundamental Research State Fund of Ukraine, Project No F58/04.
Also K.A.B.   acknowledges  a partial support provided by the Helmholtz
International Center for FAIR within the framework of the LOEWE
program launched by the State of Hesse.


\end{document}